\documentclass[twocolumn]{aastex631}

\begin{document}

\title{exoALMA. XXIV. Formaldehyde Emission in Protoplanetary Disks of exoALMA Compared with Their Properties and Dynamical State.}

\correspondingauthor{Felipe Alarc\'on}
\email{felipe.alarcon@unimi.it}

\author[0000-0002-2692-7862]{Felipe Alarc\'on }
\affiliation{Dipartimento di Fisica, Università degli Studi di Milano, Via Celoria 16, 20133 Milano, Italy}

\author[0000-0003-4689-2684]{Stefano Facchini}
\affiliation{Dipartimento di Fisica, Università degli Studi di Milano, Via Celoria 16, 20133 Milano, Italy}

\author[0000-0002-8623-9703]{Leon Trapman}
\affiliation{Department of Astronomy, University of Wisconsin-Madison, 475 N Charter St, Madison, WI 53706, USA}

\author[0000-0003-2045-2154]{Pietro Curone}
\affiliation{Departamento de Astronomía, Universidad de Chile, Camino El Observatorio 1515, Las Condes, Santiago, Chile}

\author[0009-0001-2795-8468]{Luna Rampinelli}
\affiliation{Dipartimento di Fisica, Università degli Studi di Milano, Via Celoria 16, 20133 Milano, Italy}

\author[0000-0003-2253-2270]{Sean M. Andrews}
\affiliation{Center for Astrophysics — Harvard \& Smithsonian, Cambridge, MA 02138, USA}

\author[0000-0001-7258-770X]{Jaehan Bae}
\affiliation{Department of Astronomy, University of Florida, Gainesville, FL 32611, USA}

\author[0000-0001-6378-7873]{Marcelo Barraza-Alfaro}
\affiliation{Department of Earth, Atmospheric, and Planetary Sciences, Massachusetts Institute of Technology, Cambridge, MA 02139, USA}

\author[0000-0002-7695-7605]{Myriam Benisty}
\affiliation{Université Côte d’Azur, Observatoire de la Côte d’Azur, CNRS, Laboratoire Lagrange, France}
\affiliation{Max-Planck Institute for Astronomy (MPIA), Königstuhl 17, 69117 Heidelberg, Germany}

\author[0000-0002-5503-5476]{Maria Galloway-Sprietsma}
\affiliation{Department of Astronomy, University of Florida, Gainesville, FL 32611, USA}

\author[0000-0002-8138-0425]{Cassandra Hall}
\affiliation{Department of Physics and Astronomy, The University of Georgia, Athens, GA 30602, USA}
\affiliation{Center for Simulational Physics, The University of Georgia, Athens, GA 30602, USA}
\affiliation{Institute for Artificial Intelligence, The University of Georgia, Athens, GA 30602, USA}

\author[0000-0003-1008-1142]{John D. Ilee}
\affiliation{School of Physics and Astronomy, University of Leeds, Leeds, UK, LS2 9JT}

\author[0000-0002-2357-7692]{Giuseppe Lodato}
\affiliation{Dipartimento di Fisica, Università degli Studi di Milano, Via Celoria 16, 20133 Milano, Italy}

\author[0000-0001-5907-5179]{Christophe Pinte}
\affiliation{Université Grenoble Alpes, CNRS, IPAG, 38000 Grenoble, France}
\affiliation{School of Physics and Astronomy, Monash University, VIC 3800, Australia}

\author[0000-0002-0491-143X]{Jochen Stadler}
\affiliation{Université Côte d’Azur, Observatoire de la Côte d’Azur, CNRS, Laboratoire Lagrange, France}

\author[0000-0003-1534-5186]{Richard Teague}
\affiliation{Department of Earth, Atmospheric, and Planetary Sciences, Massachusetts Institute of Technology, Cambridge, MA 02139, USA}

\author[0000-0003-1526-7587]{David J. Wilner}
\affiliation{Center for Astrophysics — Harvard \& Smithsonian, Cambridge, MA 02138, USA}

\author[0000-0002-0661-7517]{Ke Zhang}
\affiliation{Department of Astronomy, University of Wisconsin-Madison, 475 N Charter St, Madison, WI 53706, USA}



\begin{abstract}

The presence of asymmetries and substructures in protoplanetary disks, revealed by both dust and gas emission, highlights the potential interplay and the broader connection between chemistry and dynamics in disk evolution. We explore multiple relationships using the nonparametric Kendall-$\tau$ correlation to examine formaldehyde (H$_2$CO) emission with relation to stellar and disk properties for a subset of disks from the exoALMA sample. We also retrieve the H$_2$CO column density and excitation temperature using four transitions, measured in radial bins of 100 au, and quantify the level of asymmetry in the resolved peak intensity of the H$_2$CO emission. From our correlation analysis, we find no correlations with sufficient statistical significance. However, we identify tentative relationships that can be tested with larger samples. In particular, we report a proposed correlation ($2.1\sigma$) between stellar effective temperature and the formaldehyde excitation conditions, suggesting that, to first order, the central star dominates the nature of the H$_2$CO emission over possible dynamical asymmetries traced by dust. Although a correlation with the stellar luminosity was also expected, a larger sample is required to confirm or refute this trend. A possible correlation with spectral type, together with the broad range of H$_2$CO excitation temperatures within the inner 100 au of the studied disks, hint at possible multiple chemical formation pathways for H$_2$CO, including both gas-phase reactions and ice-surface chemistry on dust grains.
\end{abstract}


\section{Introduction}

Protoplanetary disks, as the birthplaces of planets, are frequently studied because they offer key insights into planet formation. Understanding the footprint of planet-disk interactions has led to deep and very accurate studies of disk dynamics, both observationally and theoretically. From an observational point of view, the exoALMA Large Program (LP) targeted 15 disks using the Atacama Large Millimeter Array (ALMA) with exquisite spectral ($\sim 27$ m s$^{-1}$) and spatial resolution ($\sim$ 0\farcs15), providing a precise characterization of the dynamical and physical structure of these disks \citep{exoALMAI_Teague}. From the theoretical point of view, the community is now able to predict the expected signatures either from the presence of planets or produced by a diversity of dynamical instabilities \citep[e.g., in ][]{Lesur..PPVII, Pinte..2023, exoALMAXVI_Barraza}. Despite progress in the dynamical studies of protoplanetary disks, the complex interplay between the different dynamical structures and the chemical evolution still has multiple unknowns \citep[e.g.,][]{Zagaria2025}.

To begin exploring the connection between the dynamical state of disks and gaseous emission, one way forward is to start investigating potential correlations between chemical molecular tracers  and other disk properties; the exoALMA sample provides an exquisite dataset to start this exploration. Among the subset of bright millimeter-wavelength emission lines, formaldehyde (H$_2$CO) stands out as a gas temperature tracer \citep{Mangum..93} given its low asymmetry. Thus, it behaves almost as a symmetric top molecule with transitions mostly dominated by collisions. In addition, many of its multiple transitions are  simultaneously observable with the ALMA interferometer within the same spectral window. It is also of particular interest as a prebiotic molecule, as it is involved in the formation of more complex organic species. In particular, formaldehyde plays an important role in the chemical evolution of disks as a key intermediate step in the formation of complex organic molecules (COMs) in warm gas \citep{Snyder..96} -- precursors of subsequent amino acids and biotic molecules \citep{Garrod..2008} -- through consecutive hydrogenation reactions. It also plays a significant role in cold grain-surface reactions, for example, leading to the production of methanol \citep{Fedoseev..2017}. Constraining the dominant formation pathway can provide insight into the development of chemical complexity and the influence of dynamical evolution on this process.

Formaldehyde can have more than one origin, as it has multiple formation pathways. One of them is through cold ($\sim$ 20 K) hydrogenation reactions of CO ice that take place on the dust grain surfaces \citep{Tielens..Hagen..1982, Cuppen..2009} that release gas-phase H$_2$CO mostly through nonthermal desorption mechanisms as the dust temperature is below the formaldehyde's sublimation temperature. Another path is through neutral-neutral gas-phase reactions that involve the oxidation of methyl (CH$_3$), leading to the production of H$_2$CO \citep{Fockenberg..2002, Atkinson..2006, Ramal-olmedo..2021}. Previous research in unresolved disks showed that formaldehyde's emission in protoplanetary disks has a low excitation temperature consistent with the ice formation path \citep{Qi..2013}. However, other studies show a possible stronger production of formaldehyde through gas-phase chemistry instead of the ice formation mechanism \citep[e.g. ][]{Carney..2019, van..Scheltinga..21}, or a combination of both \citep{Oberg..2017}. This trend is supported by an ALMA survey conducted by \cite{Pegues..20}, which concluded that the formation of H$_2$CO is a mixture of the gas-phase and grain-surface formation pathways. The fact that H$_2$CO can originate from both ice sublimation and gas-phase formation raises an intriguing question about the influence of millimeter dust structures on its abundance and on the subsequent development of chemical complexity within disks.

One of the results of the recent exoALMA survey is the estimate of the nonaxisymmetric index (NAI) in the millimeter dust continuum emission of the targeted disks \citep{exoALMAIV_Curone}. This metric attempts to quantify the level of asymmetry in the protoplanetary disk millimetric dust distribution, which is expected to be connected to the dynamical state of the gas. A dynamically disturbed disk should also show imprints in thermochemical properties such as temperature and gas density, which follow these perturbations \citep[e.g., ][]{Molyarova..2021}. It is then expected that a highly asymmetric disk will show morphological features in molecular emission, either through the excitation of specific lines or localized chemical changes that affect  the line emission.  Thus, the NAI can be compared with the emission of different molecular tracers to understand the connection between pebbles and gas, and also between the disk dynamics and chemistry through an empirical, purely observational approach.

H$_2$CO, compared to other bright tracers, e.g., CO, has the advantage of having multiple observable transitions spanning a broad range of energy levels in a single setup. Therefore, it can act as a gas thermometer and tracer of the development of chemical complexity. Thus, the properties of the H$_2$CO emission may be correlated with the dynamical state of the hosting disks, which can be indirectly inferred by the NAI of the continuum emission. In this paper, we thus present new observations of four H$_2$CO transitions of a subset of the exoALMA sample, with the aim to (1) search for correlations of the observed emission with stellar and disk properties; (2) shed light on the origin of gas-phase H$_2$CO; (3) specifically investigate whether the emission morphology of H$_2$CO is correlated to the continuum emission \citep[as for the IRS~48 disk,][]{van_der_Marel2021,Temmink..IRS}, indirectly hinting at a potential connection between the chemical and dynamical state of disks.

This paper is organized as follows. We detail the sample and data used together with the methodology for flux extraction and retrieval of physical properties in Section \ref{Sect: Methods}. Then, we present the results of our retrieval and comparison with the symmetry state of the targeted disks in Section \ref{Sect: Results}. We discuss the implications of our results in Sect. \ref{Sect: Discussion} and finally summarize our findings in Section \ref{Sect: Summary}.

\section{Methodology}\label{Sect: Methods}

\subsection{Sample and Data Reduction}

The observed formaldehyde lines are covered by the
2022.1.00485.S and 2023.1.00334.S programs (PI: L. Trapman) in a single spectral window (see Table \ref{Table: Lines}). The range in upper state energies of those lines, from 35 K to 141 K, provides a span broad enough to perform an effective rotation diagram analysis of the H$_2$CO gas in the observed disks.

 \begin{table}
 \begin{center}
      \caption{Formaldehyde lines in the targeted disks. The Einstein coefficients, upper level energies and the respective degeneracies for each transition are obtained from the Cologne Database for Molecular Spectroscopy \citep{CDMS}.}
         \label{Table: Lines}
         \begin{tabular}{ccccc}
            \hline
            \noalign{\smallskip}
            Line       &  Frequency & log(A$_{ul}$)  & E$_u$ & g$_u$ \\
            H$_2$CO      &  (GHz) &  (s$^{-1}$) & (K) &  \\
            \noalign{\smallskip}
            \hline
            \noalign{\smallskip}
            4(0,4)-3(0,3) & 290.623405 & -3.16 & 34.90403 & 9      \\
            4(2,3)-3(2,2) & 291.237766 & -3.28 & 82.07088 & 9    \\
            4(3,2)-3(3,1) & 291.380442 & -3.52 & 140.93833 & 27      \\
            4(3,1)-3(3,0) & 291.384362 & -3.52 & 140.93866 & 27    \\
            \noalign{\smallskip}
            \hline
         \end{tabular}
 \end{center}
   \end{table}

Our sample is formed by the exoALMA sources that have H$_2$CO observations of multiple lines, allowing us to retrieve the physical condition of the formaldehyde gas and compare their asymmetry with the NAI from the dust continuum emission. The ten sources successfully observed by the 2023.1.00334.S program  \citep{ExoALMAXIII_Trapman} are a subsample of the exoALMA targets. We additionally include Elias 2-27, which was observed in the 2022.1.00485.S program with the same spectral setup covering the formaldehyde emission. Elias 2-27 is a source of interest as it has precise dust continuum observations featuring grand-design spirals,  characterizing it as a dynamically active object \citep{Perez..2016,DSHARP..III, Paneque..21}. Therefore, it can provide useful insights into the exploration of the link between dynamics and chemistry.  All the analyzed sources are listed in Table \ref{Table: Disks}. The calibration of the dataset is the same as that performed by \citet{ExoALMAXIII_Trapman}. 

\setlength{\tabcolsep}{4pt}
\begin{table*}
\begin{center}
      \caption{Disk and stellar properties of the targeted sample.}
         \label{Table: Disks}   
         \begin{tabular}{lcccccccccccr}
 \tableline
Disk &  PA & Inclination  & Distance  &  V$_{\mathrm{lsr}}$& $M$  & $\dot{M}$ & $M_{\mathrm{dust}}$ & $L_*$ & T$_{\mathrm{eff}}$ & NAI & NAI$_{\mathrm{H_2CO}}$ & References \\
            &  (deg) & (deg)  & (pc)  &  (km s$^{-1}$) & $(M_{\odot})$ &  $(M_{\odot}/yr)$ & $(M_{\oplus})$ & $(L_{\odot})$ & (K) &  & & \\
            \tableline
            AA Tau &  273 & -59 & 135 & 6.50 & 0.79 & 10$^{-8.4}$ & 27.8 &1.1 & 4350 & 0.12 & 0.535 & 1-5 \\
            CQ Tau &  235 & 36 & 149 & 6.19 & 1.40 & 10$^{-7.0}$ & 44.4 & 10.0 & 6890 & 0.111 & 0.452 & 1-4,6 \\
            HD 34282 &  117 & -58 & 309 & -2.33 & 1.62 & 10$^{-7.7}$ & 109.4 & 10.8 & 9520 & 0.114 & 0.591 & 1-4,7\\
            HD 135344B &  243 & -16 & 135 & 7.09 & 1.61 & 10$^{-7.4}$ & 28.1 & 6.7 & 6440 & 0.405 &0.480 & 1-4,7\\
            HD 143006 &  168 & -17 & 167 & 7.72 & 1.56 & 10$^{-8.1}$ & 24.6 & 3.8 & 5620 & 0.215 & 0.151 & 1-4,8\\
            MWC 758 &  240   &   19 & 156 & 5.89 & 1.40 & 10$^{-7.7}$ & 19.1 & 14.0 & 7850 & 0.429 & 0.560 & 1-4,9 \\
            PDS 66 &  189   &   -32 & 98 & 3.96 & 1.28 & 10$^{-9.9}$ & 30.4 & 1.2 & 4189 & 0.014 &0.261& 1-4,10\\
            RX J1615 &  325   &   46 & 156 & 4.75 & 1.14 & 10$^{-8.5}$ & 312.3 & 0.6 & 4350 & 0.038 & 0.196& 1-4,11 \\
            RX J1842 &  206   &   39 & 151 & 5.94 & 1.07 & 10$^{-8.8}$ & 36.6 & 0.8 & 4780 & 0.074 &0.380& 1-4,11\\
            RX J1852 &  117   &   -33 & 147 & 5.47 & 1.03 & 10$^{-8.7}$ & 33.2 & 0.6 & 4780 & 0.024 & 0.424& 1-4,11\\
            Elias 2-27 &  56.0   &   56.0 & 116 & 2.0 & 0.49 & 10$^{-7.2}$ & 103.9 & 0.92 & 3850 & 0.108 & 0.266& 1,4,12  \\

            \tableline
         \end{tabular}
\end{center}
    \tablecomments{With the exception of Elias 2-27, the PA, inclination, $M$, and V$_{\mathrm{lsr}}$ are the ones listed in \cite{Izquierdo..2025}. The disk dust mass, effective temperature and luminosity are the ones listed in \cite{ExoALMAXIII_Trapman}. The stellar distance is measured by the Gaia collaboration \citep{Gaia}, while accretion rates are retrieved from the literature. The NAI of the exoALMA sources was calculated by \cite{exoALMAIV_Curone}, while the NAI of Elias 2-27 is calculated in the current work with the same method. The NAI$_{\rm H_2CO}$ for all the sources is calculated in this work.}
     \tablerefs{(1) \cite{Gaia}. (2) \cite{Izquierdo..2025}. (3) \cite{exoALMAIV_Curone}. (4) \cite{ExoALMAXIII_Trapman}. (5) \cite{Bouvier..2013}. (6) \cite{Donehew..2011}. (7) \cite{Fairlamb..2015}. (8) \cite{Rigliaco..2015}. (9) \cite{Boehler..2018}. (10) \cite{Ingleby..2013}. (11) \cite{Manara..2014}. (12) \cite{NAtta..2006}}
\end{table*}

We imaged the cubes using the CASA software \citep{CASA} with \textit{natural} weighting to maximize the signal-to-noise ratio (SNR) for the H$_2$CO emission, trading the SNR with the spatial resolution. The resulting data cubes have a native spectral resolution of $\sim$ 1.2 km s$^{-1}$ at the frequency of the H$_2$CO lines listed in Table \ref{Table: Lines}, and a spatial resolution ranging between 0$\farcs$3 and 0$\farcs$4. This spatial resolution was good enough to sample all disks with at least two radial bins of 100 au each, which allowed us to compare the retrieved physical quantities in the inner regions ($<$100 au) under a uniform physical criterion, and beyond the CO snowline of the targeted disks ($>$100 au). The extent of these radial bins is arbitrary, but it secures us at least one spatial beam for the most distant target. It also matches the expected resolution of larger disk samples, such as the observations performed by the DEC/O ALMA 2022.1.00875.L Large Program. The emission of a few disks extends beyond 200 au; however, we choose to compare equivalent physical scales, even considering biases associated with it, since thresholds for radial bins based on the extension of the emission and/or stellar luminosity lead to distinct emission areas for different transitions. Thus, given the good spatial resolution of the data, we opted not to perform the rotational diagram analysis using the integrated fluxes from the whole disk, but only within the aforementioned radial bins.

\subsection{Flux extraction and retrieval of physical parameters}

We compute integrated intensity (moment 0) maps for each line by integrating the data cubes over spectral ranges varying between 3.6 and 6 km s$^{-1}$  (three to five spectral channels). For each disk, we chose a spectral band with emission observable higher than 3$\sigma$ for the brightest H$_2$CO transition. Then, the same velocity range was applied for the other transitions.  After that, we extract the line fluxes from the integrated intensity maps, encompassing the whole disks. Since the H$_2$CO  4$_{3,2}$-3$_{3,1}$ and 4$_{3,1}$-3$_{3,0}$ lines are partly blended, present weak emission, and have almost identical quantum numbers (see Table \ref{Table: Lines}), we compute the average flux between those two lines. Then, the average flux is used for the rotation diagram analysis described below as a single data point. This approach boosts the SNR for these higher energy transitions while being able to provide additional information for our retrieval of the physical properties of the formaldehyde excitation conditions. The total fluxes and their uncertainties are reported in Table \ref{Table: Fluxes}.

\begin{table}
\begin{flushleft}
      \caption{Integrated disk fluxes of the H$_2$CO emission.}
         \label{Table: Fluxes}   
         \resizebox{0.48\textwidth}{!}{
         \begin{tabular}{lccc}
 \tableline
Disk &  F$_{\mathrm{H_2CO}, 4_{0,4}-3_{0,3}}$  & F$_{\mathrm{H_2CO}, 4_{2,3}-3_{2,2}}$  & F$_{\mathrm{H_2CO},  4_{3,2}-3_{3,1}, 4_{3,1}-3_{3,0}}$ \\
            &  (mJy km s$^{-1}$) & (mJy km s$^{-1}$) & (mJy km s$^{-1}$)  \\
            \tableline
            AA Tau &  229$\pm$27 & 31 $\pm$7 & 75$\pm$22 \\
            CQ Tau &  382$\pm$28 & 158$\pm$52 & 243$\pm$63 \\
            HD 34282 &  231$\pm$29 & 54$\pm$33 & 53$\pm$37 \\
            HD 135344B &  346$\pm$67 & 70$\pm$23 & 110$\pm$45\\
            HD 143006 &  535$\pm$49 & 88$\pm$57 & 4$\pm$25 \\
            MWC 758 &  218$\pm$14   &  69$\pm$35 & 54$\pm$41 \\
            PDS 66 &  171$\pm$20   &  23$\pm$14 & $<$25 \\
            RX J1615 &  706$\pm$36   &  27$\pm$28 & 71$\pm$24 \\
            RX J1842 &  375$\pm$16   &  20$\pm$7 & 7$\pm$42 \\
            RX J1852 &  422$\pm$40   &  96$\pm$23 & 40$\pm$26 \\
            Elias 2-27 &  498$\pm$29   &   17$\pm$10 & $<$17  \\ 
                    
            \tableline
         \end{tabular}}
\end{flushleft}
\tablecomments{Flux uncertainties do not include the uncertainty in the absolute flux calibration of the ALMA interferometer. The fourth column is the average flux between the H$_2$CO $4_{3,2}-3_{3,1} \mathrm{and}  4_{3,1}-3_{3,0}$ lines.}
\end{table}

\begin{figure*}
   \centering
   \includegraphics[width=0.495\textwidth]{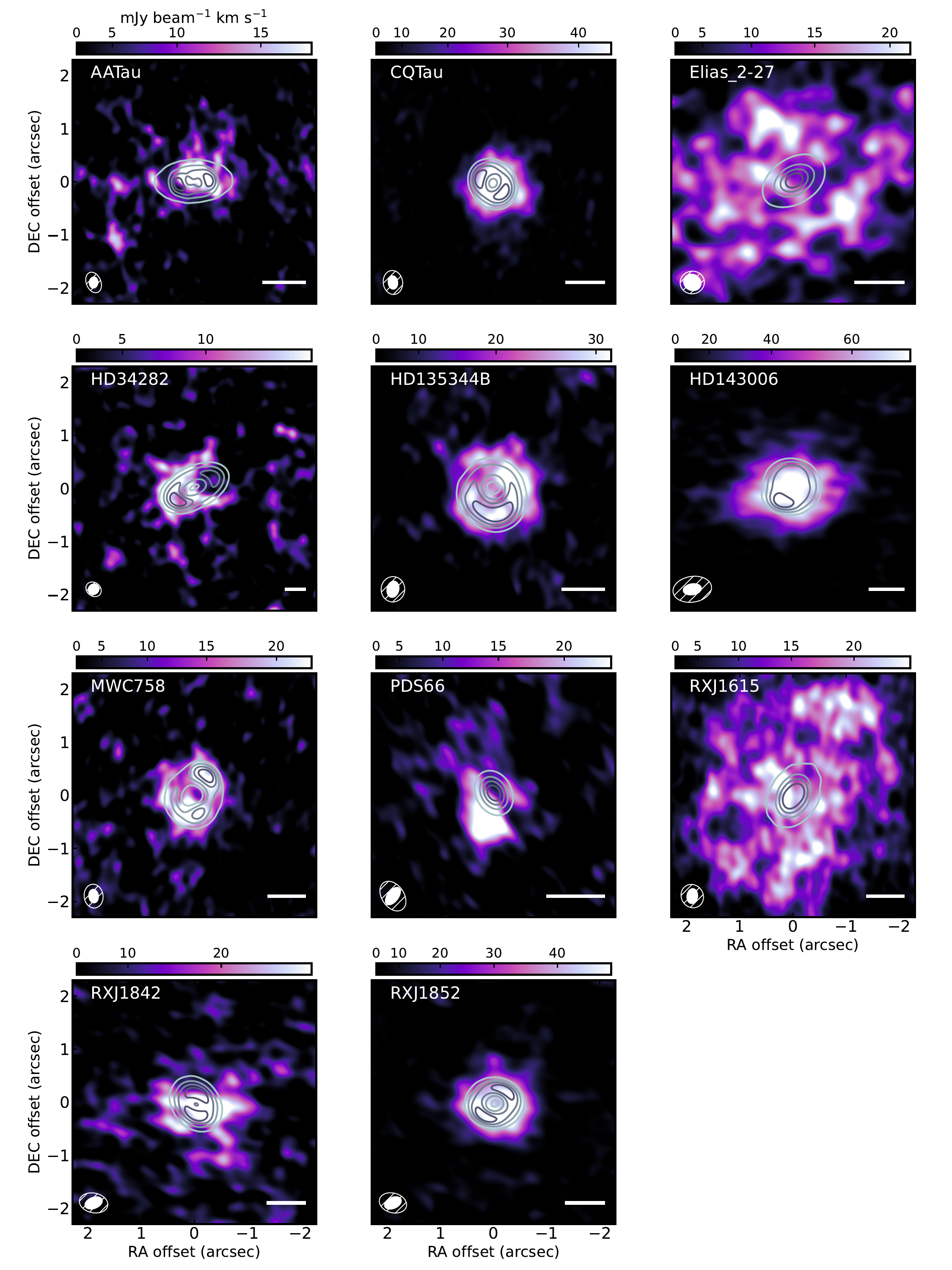}
   \includegraphics[width=0.495\textwidth]{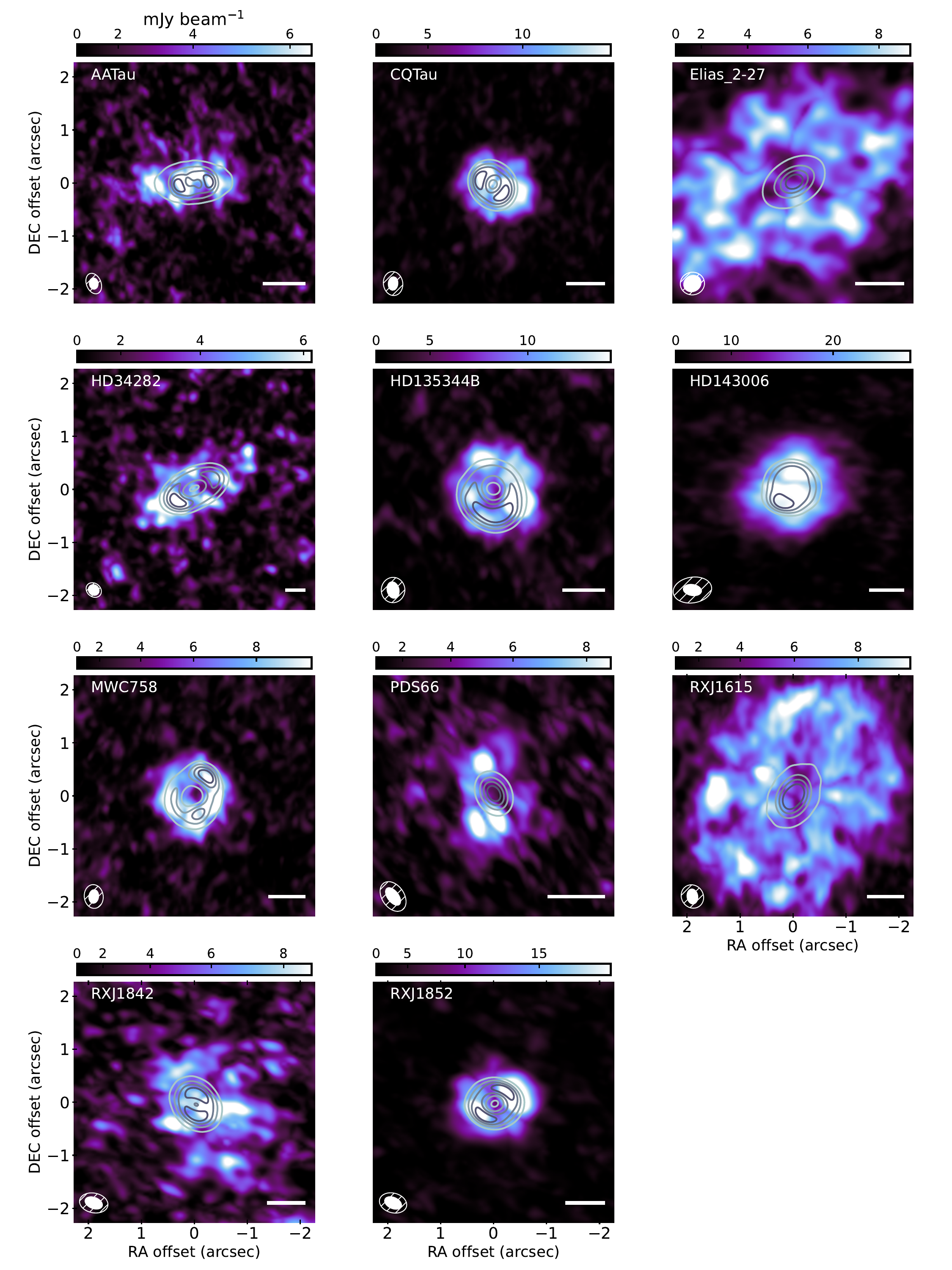}
      \caption{\textbf{Left}: integrated emission maps of the H$_2$CO 4(0,4)-3(0,3) line for the observed sources. \textbf{Right:} peak intensity emission for the same line. Continuum emission at 0.8 mm contours is overlaid to compare with the formaldehyde emission in the disks. Beams are illustrated on the bottom-right with solid white for continuum images and dashed lines for line emission. The bottom-left bar represents a respective 100 au physical scale for each source. For most disks, there are no clear correspondences between the dust continuum contours and the integrated maps.}
         \label{Fig:MOM0}
   \end{figure*}


To make a coherent comparison across the different sources, we also extract line fluxes in two distinct elliptical apertures for the four transitions on each disk physical frame, considering projection effects. The first aperture corresponds to the inner 100 au, while the second aperture spans from 100 to 200 au. While in some cases the emission extends beyond the 200 au radius, we keep the outer radius fixed to maintain the comparison between disks consistent. The fluxes extracted in these apertures are reported in Tables~\ref{Table: Disks_in} and \ref{Table: Disks_out}.


Given that the H$_2$CO  4$_{3,2}$-3$_{3,1}$ and 4$_{3,1}$-3$_{3,0}$ are ortho-transitions, while H$_2$CO 4$_{0,4}$-3$_{0,3}$ and  4$_{2,3}$-3$_{3,3}$ are para-lines, we note that the combined partition function intrinsically assumes a gas-phase ortho-to-para ratio (OPR) scaling of 3.0. In general, this value is roughly in agreement with laboratory experiments at low temperatures \citep[$<$50 K,][]{Yocum..23}, and close to values reported in observations \citep[even though some cases report values $<3$,][]{Guzman..2018,van..Scheltinga..21, hernandez_vera..2024}.

After extracting the fluxes, we calculate their associated uncertainties with a bootstrapping approach following a similar method to \citet{Luna..24}. First, we extract the flux in multiple nonoverlapping elliptical apertures ($\geq$15) in emission-free regions of the integrated intensity maps. These apertures are equivalent in size and geometry to the ones used for each specific disk. We then estimate the error by calculating the standard deviation of the extracted fluxes in those apertures.  Even if we avoid sampling the edges of the field of view, we opt for a  conservative approach as it ends up overestimating the uncertainties of our measurements given the spatial correlation of the noise in interferometric observations \citep{Tsukui..22}.

Once we retrieve the line fluxes, we employ rotation diagrams to recover the physical conditions of the formaldehyde gas \citep{Goldsmith..Langer..1999}, where we assume local thermodynamic equilibrium. Assuming optically thin emission of the formaldehyde lines, the  column density $N$, and excitation temperature $T_{\mathrm{exc}}$ are linked through the following relationship:

\begin{equation}\label{Eq:rot}
    N = \frac{Q(T_{\mathrm{exc}})}{g_u}\exp{\Big(\frac{E_u}{T_{\mathrm{exc}}} \Big)}\frac{4\pi}{A_{ul}hc}\frac{S_{\nu}\Delta \nu}{\Omega},
\end{equation}

 \noindent with $Q$ the partition function, $g_u$ the degeneracy of the upper state, E$_u$ the upper state energy, $A_{ul}$ the Einstein coefficient of the respective transition, $\Omega$ the solid angle of the integrated area, and $S_{\nu}\Delta \nu $ the line flux.  The partition function is calculated by interpolating between tabulated values from the Cologne Database for Molecular Spectroscopy \citep{CDMS}. As  mentioned previously, in the treatment of the ortho- and para-transitions, we assume common physical conditions and a common partition function considering a fixed OPR.

Using the relationship described above, we derive the column density and excitation temperature in both the inner (0-100 au) and outer (100-200 au) disk regions. We recover the physical parameters and their respective uncertainties by employing a Markov Chain Monte Carlo (MCMC) fitting as described in \cite{Luna..24} using the \texttt{emcee} package \citep{emcee}. We apply a uniform prior for the excitation temperature ranging between 10 and 200 K, and a uniform prior on the log scale of the formaldehyde column density between 10$^{10}$ and 10$^{14}$ cm$^{-2}$. For each disk, the fitting is performed using 256 walkers, and 5000 steps after a burn-in of 1000 steps. The corner plots for each disk are illustrated in Appendix \ref{Appendix: rot}.

\subsection{Correlation Analysis}

Once the physical parameters of the formaldehyde emission are retrieved, we cross-correlate them with several disk and stellar parameters, such as stellar mass, disk mass (gas and dust), stellar luminosity, and accretion rate. The goal is to search for indications of causal connections between these properties and the chemistry and excitation conditions for H$_2$CO. We also look for possible trends that may have a more (less) significant effect in the inner (0-100 au) or outer (100-200 au) regions of these disks. In addition, we cross-compare the respective NAIs for the continuum and gas emission with the disk and physical parameters to look for correlations that can shed light on the link between the dynamical state of these disks and their gas emission. We opted to use the nonparametric Kendall-$\tau$ correlation given the size of our disk sample \citep{SHIEH199817}. When observational uncertainties were present, weights were applied in the calculation of the correlation coefficients. The respective weights for each parameter are: $w_i = (1 + (\sigma_i/X_i))^{-1}$, with $\sigma_i$ the uncertainty associated with the measurement $X_i$. The statistical significance of the correlations is assessed by calculating the $z$-scores corresponding to each Kendall-$\tau$ correlation coefficient. Similarly to $p$-values, the $z$-scores indicate the significance of a correlation by expressing how many standard deviations the observed coefficient deviates from the expected value under the null hypothesis.

\subsection{Calculation of the NAI}

Given the intrinsic three-dimensional nature of gas emission in protoplanetary disks, employing the same definition of an NAI as defined for the continuum emission \citep{exoALMAIV_Curone} would not be appropriate. Even when accounting for the elevated emission of gas lines (which is not always possible for low SNR data, as is the case for the data in this paper), the deprojection of an emission surface of a dataset with a finite angular resolution would naturally introduce deviations from axisymmetry. A perfectly symmetric disk would, however, preserve a mirror-symmetry against the minor axis of the disk, irrespective of an elevated emission surface and finite beam size. Therefore, we calculate the molecular non-axisymmetric index on the intensity peak maps after folding them against the minor axis of the disk. We provide a more detailed discussion about our metric choice and its robustness in Appendix \ref{App:NAI}. The particular metric that we use as NAI$_{\mathrm{H_2CO}}$ is

\begin{equation}\label{Eq: NAI}
    \mathrm{NAI}_{\mathrm{mol}} = \frac{\sum_{i,j} \vert I_{i,j} - I_{-i,j}\vert }{\sum_{i,j} \vert I_{ij}\vert},
\end{equation}

\noindent  where $I_{ij}$ is the peak intensity value at pixel $i,j$, and $I_{-i,j}$ is the mirroring pixel with respect to the minor axis of the disk. The difference with respect to the metric used by \cite{exoALMAIV_Curone} is that we do not add up the residuals from an azimuthally averaged profile, but instead we compare one half of the disk to its mirroring half with respect to the minor axis. We employ a 4$\sigma$ threshold on pixel pairs from the opposite sides of the disk with respect to the minor axis. If both pixels are below the given threshold, the pair is not included in the metric calculation. A 4$\sigma$ threshold was chosen as it minimizes the contribution  of imaging artifacts and very low signal-to-noise regions of the maps, while keeping a low number of pixels discarded. Depending on the sensitivity of the data, a stricter threshold will discard more pixels that may be tracing meaningful asymmetries, while a threshold that is too low will artificially boost the measured metric by including the noise-induced asymmetries. This choice was also tested over the CO line emission cubes of PDS 66 using the exoALMA data, producing consistently low values in the most symmetric disk in the sample for CO and continuum emission. We calculate the NAI$_{\mathrm{H_2CO}}$ using only the H$_2$CO 4$_{0,4}$-3$_{0,3}$ transition, which is the brightest and most extended. For the other transitions, the sensitivity threshold requires deeper observations; the number of points considered for the index would not be representative. To be consistent with our chosen regions, we calculate the NAI$_{\mathrm{H_2CO}}$ within 200 au for each disk.

\begin{figure*}
   \centering
   \includegraphics[width=0.49\textwidth]{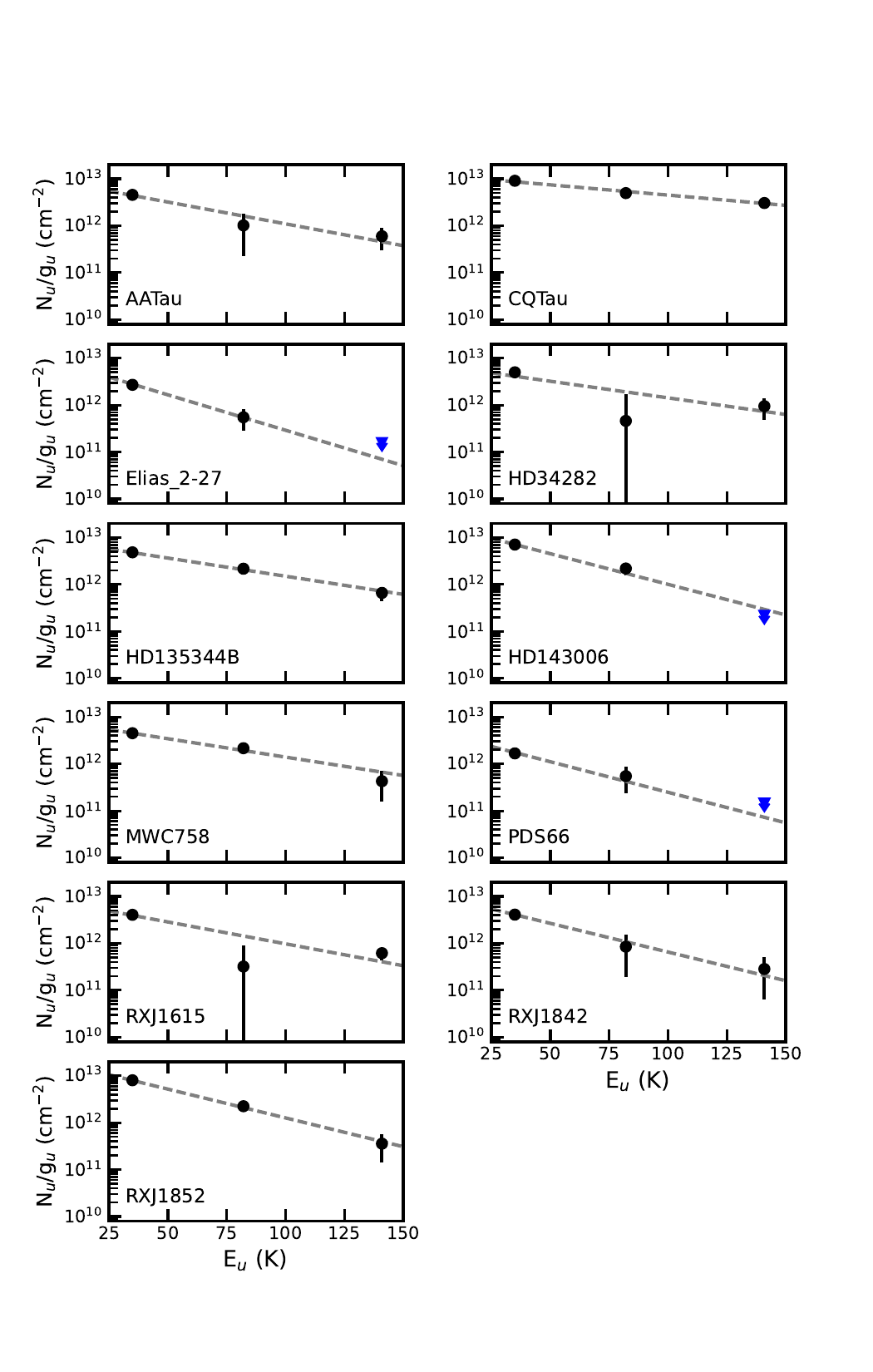}
   \includegraphics[width=0.49\textwidth]{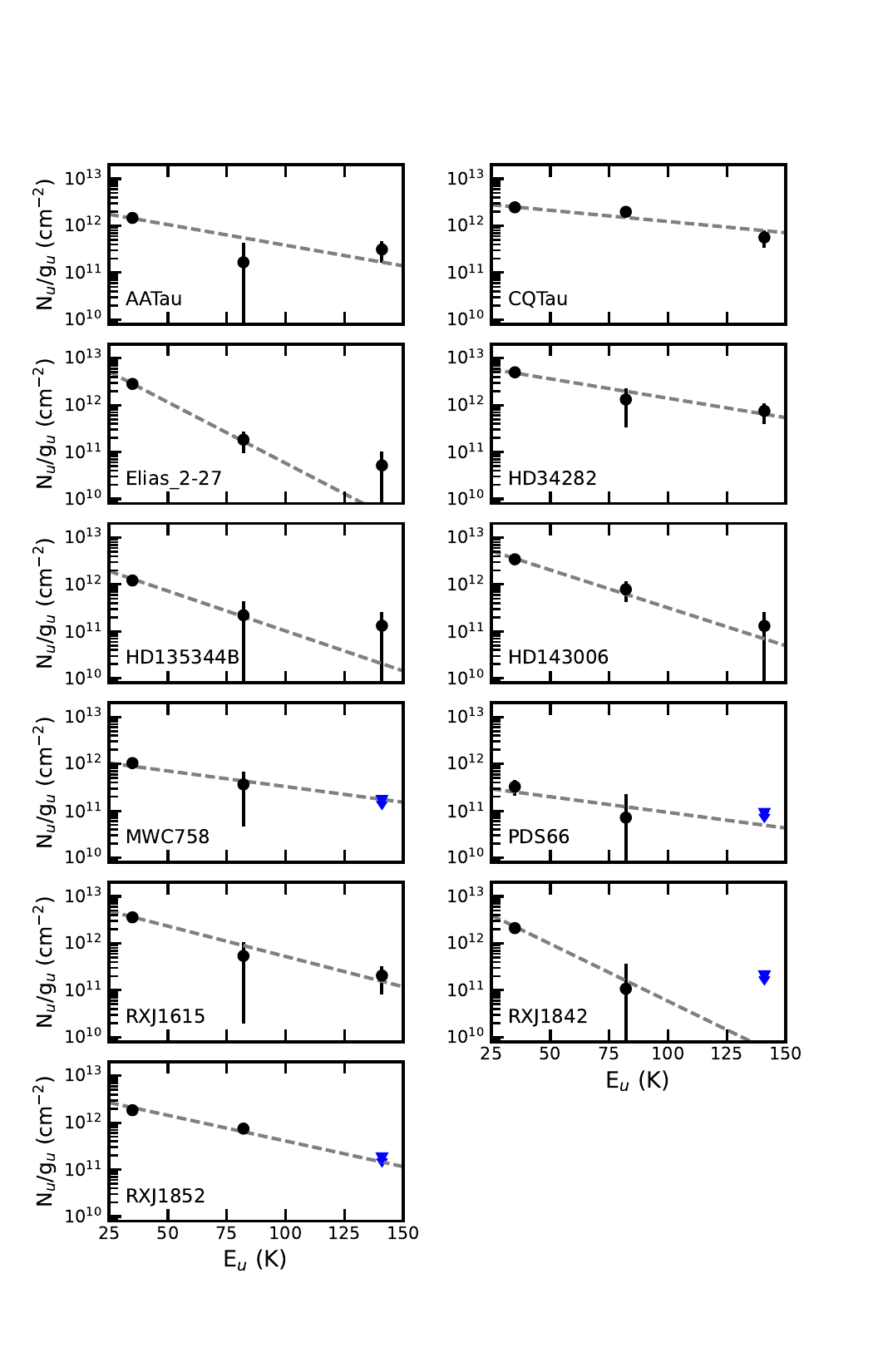}
      \caption{Rotational diagrams of the 11 targeted disks for the inner 100 au (left) and 100-200 au region (right). Blue triangles represent upper limits for undetected transitions, and the dotted lines show the best MCMC fit. In AA Tau, HD 34282, MWC 758, and RX J1615 it is possible that there is more than one emitting component.
              }
         \label{Fig:Rot_diagrams}
          \label{Fig:Rot_diagrams_out}
\end{figure*}

\section{Results}\label{Sect: Results}

\label{Sect: Results}

\begingroup
\renewcommand{\arraystretch}{1.3}
   \begin{table*}
      \caption{Targeted sources and retrieved physical parameters of the H$_2$CO emission for the inner 100 au of each disk.}
         \label{Table: Disks_in}
   $$ 
         \begin{array}{lccccc}
            \hline
            \noalign{\smallskip}
            \mathrm{Disk} &  F_{\mathrm{H_2CO}, 4_{0,4}-3_{0,3}}  & F_{\mathrm{H_2CO}, 4_{2,3}-3_{2,2}}  & F_{\mathrm{H_2CO},  4_{3,2}-3_{3,1}, 4_{3,1}-3_{3,0}}  &   \mathrm{T_{exc, in}} & \mathrm{log(N_{H_2CO, in})}\\
            &  \mathrm{(mJy\ km\ s^{-1})} & \mathrm{(mJy\ km\ s^{-1})}   & \mathrm{(mJy\ km\ s^{-1})}  & \mathrm{(K)} & \mathrm{(cm^{-2})} \\
            \noalign{\smallskip}
            \hline
            \noalign{\smallskip}
            \mathrm{AA \ Tau} & 93.0 \pm 11.6 & 15.7 \pm 12.2 & 16.15 \pm 8.17 & 46.8^{+15.3}_{-15.2} & 13.22^{+0.10}_{-0.09} \\
            \mathrm{CQ\ Tau} &  182.2 \pm 14.2 & 75.1 \pm 7.8 & 80.95 \pm 7.39 & 99.1^{+13.3}_{-10.7} & 13.83^{+0.04}_{-0.04}\\
            \mathrm{HD\ 34282} &   20.1 \pm 4.8 &  <3.7 & 5.05 \pm 2.04 & 61.6^{+36.7}_{-28.0} & 13.30^{+0.15}_{-0.16} \\
            \mathrm{HD\ 135344B} & 185.3 \pm 11.7  & 62.6 \pm 9.6 & 33.5 \pm 11.18 & 56.0^{+7.9}_{-7.2}  & 13.33^{+0.05}_{-0.05} \\
            \mathrm{HD\ 143006} &  176.4 \pm 10.5 & 41.1 \pm 12.0 &  <7.02 & 33.2^{+5.5}_{-5.9}  & 13.34^{+0.04}_{-0.04} \\
            \mathrm{MWC\ 758} &  127.0 \pm 13.0 & 45.9 \pm 7.0 & 15.95 \pm 10.05 & 55.4^{+9.9}_{-8.4}  & 13.29^{+0.06}_{-0.05} \\
            \mathrm{PDS\ 66} &  107.2 \pm 8.7 & 26.5 \pm 15.0 & <12.4 & 33.3^{+13.6}_{-13.7}  & 12.73^{+0.09}_{-0.06} \\
            \mathrm{RX\ J1615} &  82.1 \pm 6.2 & <8.7 & 16.55 \pm 5.06 & 46.3^{+10.7}_{-12.1}  & 13.17^{+0.07}_{-0.07} \\
            \mathrm{RX\ J1842} &  100.8 \pm 15.3 & 15.8 \pm 12.2 & 9.25 \pm 7.14 & 35.4^{+12.55}_{-14.3}  & 13.11^{+0.09}_{-0.08}\\
            \mathrm{RX\ J1852} &  225.3 \pm 9.3 & 47.6 \pm 11.7 & 13.3 \pm 8.11 & 35.3^{+4.8}_{-5.0} & 13.40^{+0.03}_{-0.03}\\
            \mathrm{Elias\ 2-27} & 81.5 \pm 9.7 & 12.5 \pm 6.0 & <6.2 & 28.6^{+8.8}_{-9.0}  &12.91^{+0.06}_{-0.06} \\

            \noalign{\smallskip}
            \hline
         \end{array} $$
         \tablecomments{The fourth column is the average flux between the H$_2$CO $4_{3,2}-3_{3,1} \mathrm{and}  4_{3,1}-3_{3,0}$ lines.}
   \end{table*}

We show the overlay of the integrated intensity and peak intensity maps of the H$_2$CO  4$_{3,2}$-3$_{3,1}$ line and the 0.8 mm dust continuum emission from the same dataset using contours in Figure \ref{Fig:MOM0}. Even though in some particular cases there are peaks or dips of emission colocated with the presence of a dust clump or horseshoes, there is no clear apparent trend linking the presence of dust substructures and the formaldehyde emission morphology for the different disks. No clear corresponding relationships were found in the other formaldehyde transitions either (see Appendix \ref{App: Maps}).

We illustrate the resulting fits of the observational rotational diagrams for the physical temperatures in the inner 100 au and the 100-200 au regions in Figure \ref{Fig:Rot_diagrams_out}. The best-fit parameters for the inner and outer regions are reported in Tables~\ref{Table: Disks_in} and \ref{Table: Disks_out}, respectively. The values of the retrieved column densities range between 5.4 $\times 10^{12}$ cm$^{-2}$ and 6.8 $\times 10^{13}$ cm$^{-2}$, which fall within the range of the literature-collected values in \cite{Oberg..annrev}, spanning from 10$^{12}$ cm$^{-2}$ to 10$^{14}$ cm$^{-2}$. The formaldehyde excitation temperature ranges from 17 K in the 100-200 au region of Elias 2-27 to 99 K in the inner 100 au of CQ Tau. We first assess the validity of the optically thin emission assumption by calculating the line optical depths using the retrieved quantities. Even in the disk with the highest column densities, the maximum optical depth measured is $\leq$0.15. Then, we show a comparison between the retrieved physical values for the two bins in Figure \ref{Fig: Retrieved} to assess the possible radial gradient between the two different radial bins. The formaldehyde column density retrievals show a decay between the inner 100 au and the subsequent radial bin for all but two sources, while no statistically significant difference is found between the excitation temperatures of the formaldehyde gas in the two radial regions. We discuss possible causes for the column density decrease between radial bins in Section \ref{Sect: Discussion}.

\begin{figure}
   \centering
   \includegraphics[width=0.48\textwidth]{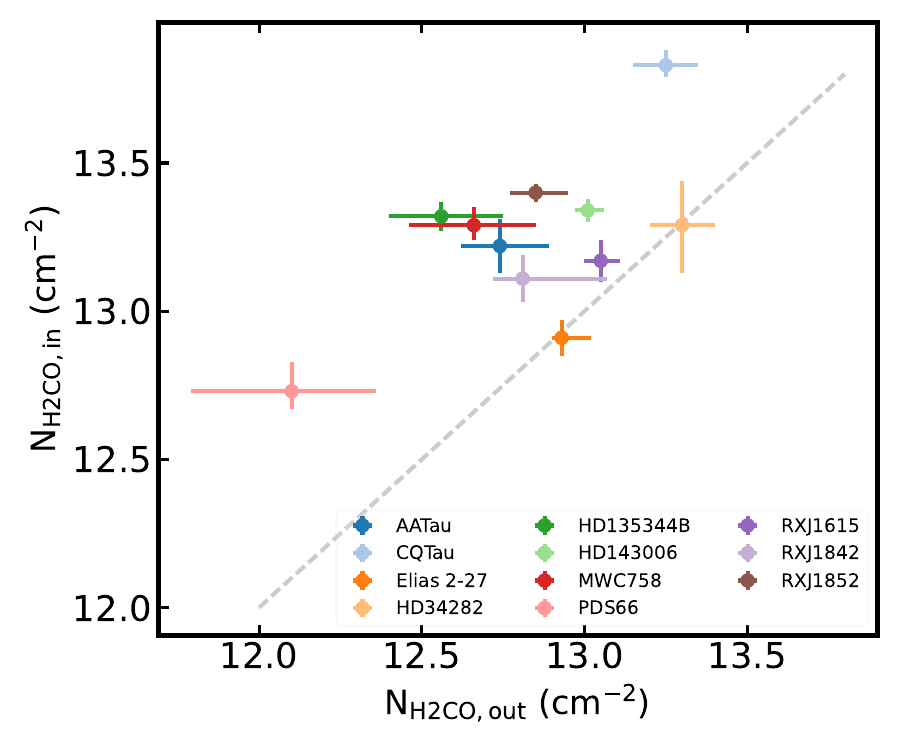}
   \includegraphics[width=0.48\textwidth]{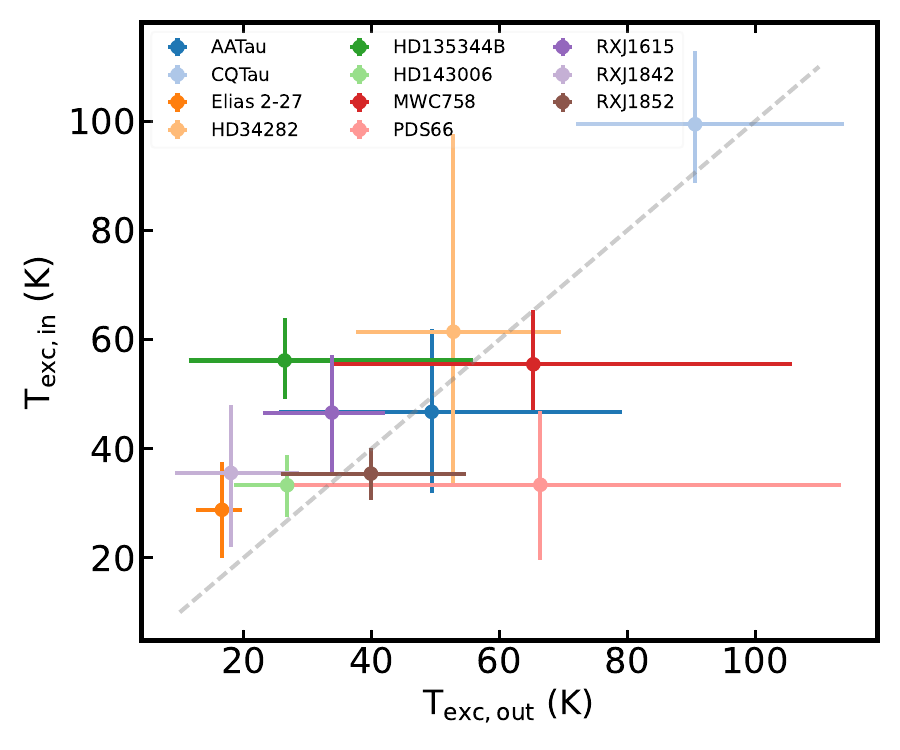}
      \caption{\textbf{Top:} comparison between formaldehyde column density in the two 100 au radial bins, with the dashed line indicating the identity. \textbf{Bottom:} excitation temperature of H$_2$CO between the inner 100 au and the successive 100-200 au radial bin. No significant difference between the two temperatures can be stated given the uncertainties due to the lower SNR of the emission in the outer 100 au.
              }
         \label{Fig: Retrieved}
   \end{figure}

\subsection{Correlation between gas physical parameters and Stellar/Disk parameters}

\begin{figure}
    \centering
    \includegraphics[width=0.65\linewidth]{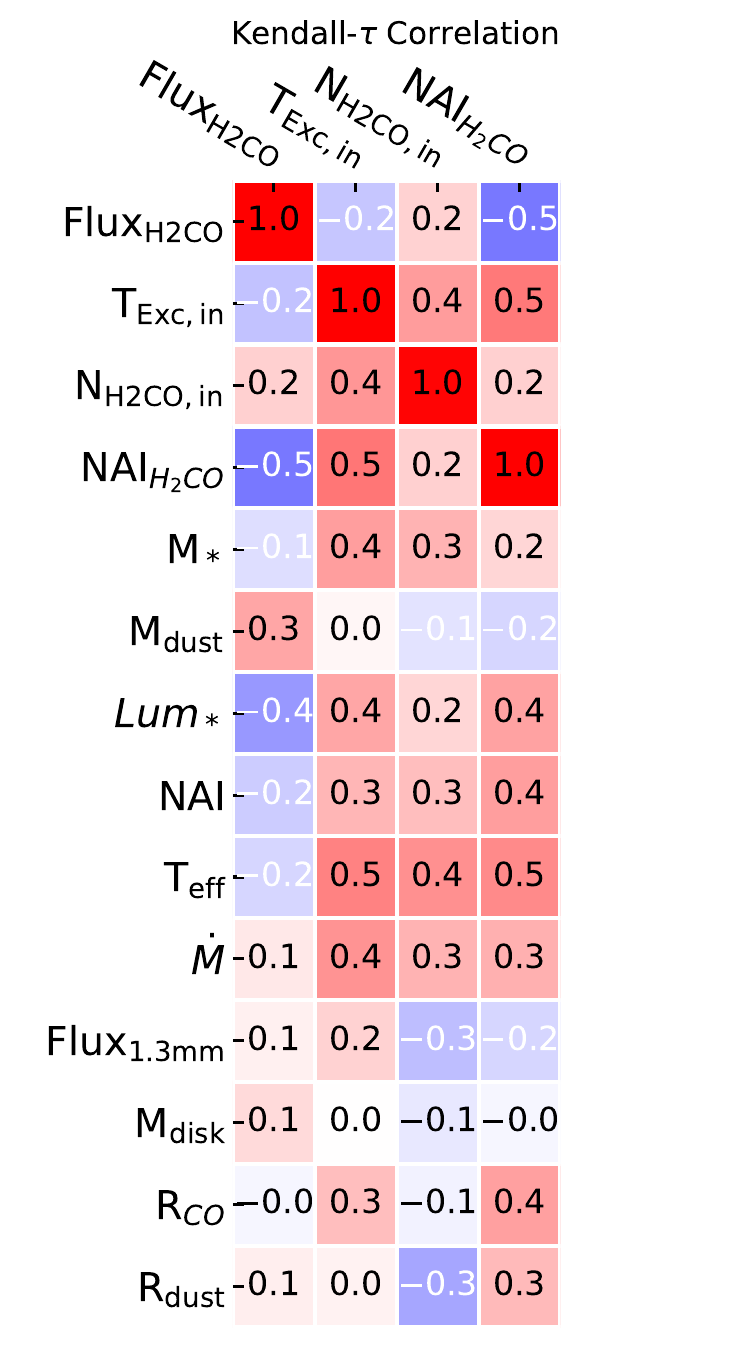}
    \includegraphics[width=0.65\linewidth]{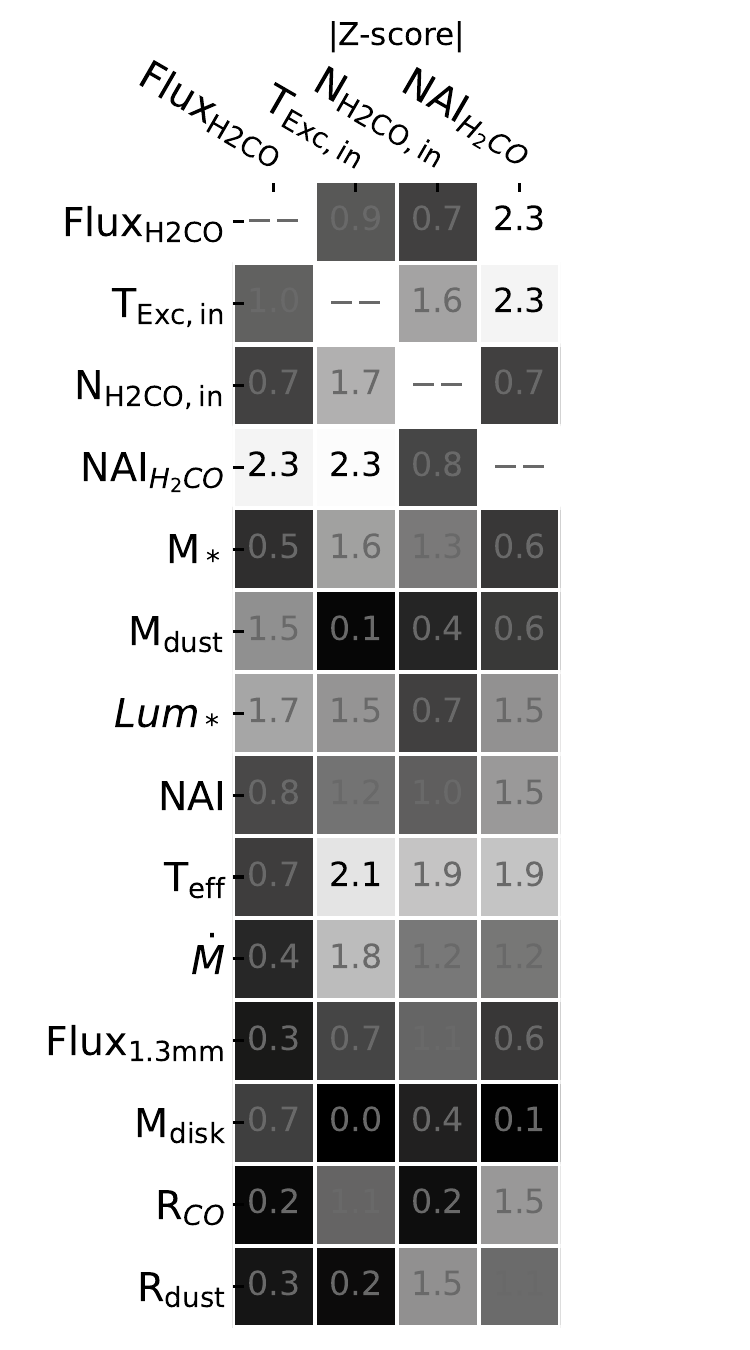}
    \caption{\textbf{Top:} weighted Kendall-$\tau$ correlation values between compared parameters/metrics, with redder more positively correlated and bluer more anticorrelated pairs. For correlation purposes, the 1.3 mm continuum flux and the line emission flux were normalized to a 100 pc distance for every disk. \textbf{Bottom:} $Z$-scores of the compared parameters and metrics representing the level of statistical significance of the weighted Kendall-$\tau$ correlation values.}
    \label{fig:zscores}
\end{figure}

Once we retrieve the physical parameters of the formaldehyde lines, we calculate the correlation matrix with  the properties listed in Table \ref{Table: Disks}. The resulting correlation matrix, together with the $z$-scores associated with the statistical significance of the correlation values, are illustrated in Figure \ref{fig:zscores}. The correlation values exceeding the 
2$\sigma$ level of significance are highlighted in white in the $z$-score matrix.


\begin{figure}
   \centering
   \includegraphics[width=0.49\textwidth]{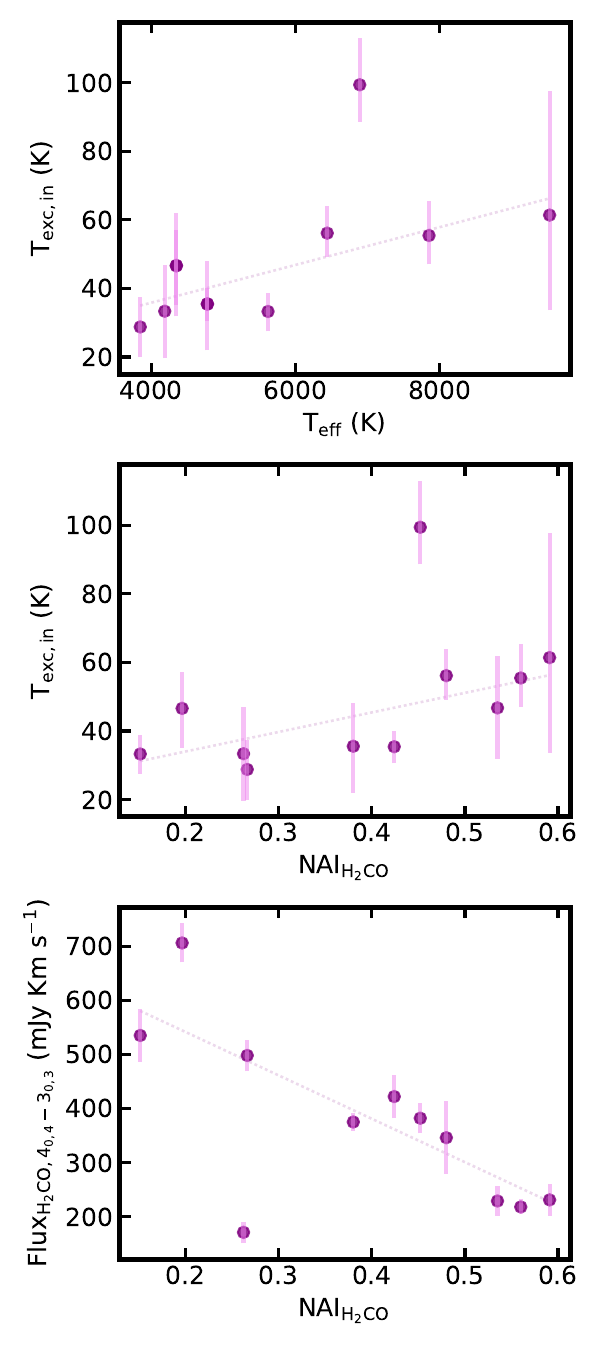}
      \caption{Subset of correlations explored with $z$-scores $\geq$2.00 between the NAI$_\mathrm{H_2CO}$, the total H$_2$CO$_{\mathrm{4(0,4)-3(0,3)}}$ line flux, and the retrieved excitation temperature of formaldehyde in the inner 100 au of the observed disks.}
         \label{Fig:corr_all}
   \end{figure}

There are expected correlations between dust continuum flux, CO disk size together with H$_2$CO flux associated with scaling effects that were not retrieved by the present dataset. A correlation analysis with the Pearson correlation index leads to a positive correlation between such indices. Nevertheless, for the Pearson correlation index to be appropriately used, a larger sample of disks is required. The current analysis using Kendall's-$\tau$ does not find  significant correlations between such parameters, although they are not discarded by the metric either.


We delve further into the origin of the correlation between $T_{\rm eff}$ and $T_{\rm exc, in}$, showing a Kendall-$\tau$ value 0.5. This relationship can be explained, to a first-order approach, through the energy balance of the dust emission with the passive heating due to the stellar irradiation. Assuming that formaldehyde traces, within reasonable uncertainties, the dust temperature, $T_{\rm dust}$, the incoming stellar flux ($F_{\rm in}$) should balance the outgoing flux ($F_{\rm out}$). In the simplistic assumption of dust particles being perfect blackbodies:


\begin{eqnarray}
F_{\rm out} & = & F_{\rm in}  \\
\sigma_{SB}T_{\rm dust}^4   & \approxeq & \frac{L_*}{4\pi r^2}  \Rightarrow T_{\rm dust} \propto L_*^{0.25} \\
\sigma_{SB}T_{\rm dust}^4  & \approxeq  &\frac{\sigma_{\rm SB} 4\pi R_*^2 T_{\rm eff}^4}{4\pi r^2}    \\
T_{\rm dust}^4  & \approxeq  &\frac{R_*^2 T_{\rm eff}^4}{r^2}  \Rightarrow T_{\rm dust} \propto T_{\rm eff},
\end{eqnarray}

\noindent where $\sigma_{\rm SB}$ is the Stefan-Boltzmann constant, $R_*$ the stellar radius, $L_*$ the stellar luminosity, $r$ a radial separation from the star. The equations above show that to first-order there is plausibility for the correlation of the formaldehyde excitation temperature with  $T_{\rm eff}$ and $L_*^{0.25}$. From the relationships shown above, a correlation with $L_*$ would also have been expected; however, it does not reach a $z$-score high enough to be statistically significant. Nevertheless, we do not discard the plausibility of a  possible correlation being found with a larger sample.

We remark a noticeable outlier case in the correlations with $T_{\rm exc, in}$ in Figure \ref{Fig:corr_all}, which is CQ Tau. CQ Tau has a high H$_2$CO excitation temperature (99 K), which separates it from the main correlation with the other disks in the sample. However, it is known that CQ Tau is a quite variable star, mainly attributed to circumstellar dust obscuration of stellar light, which may change its spectral type classification due to abrupt changes in the spectral energy distribution \citep{CQ_Tau_var}.

\subsection{NAI}

      \begin{figure}
   \centering
   \includegraphics[width=0.48\textwidth]{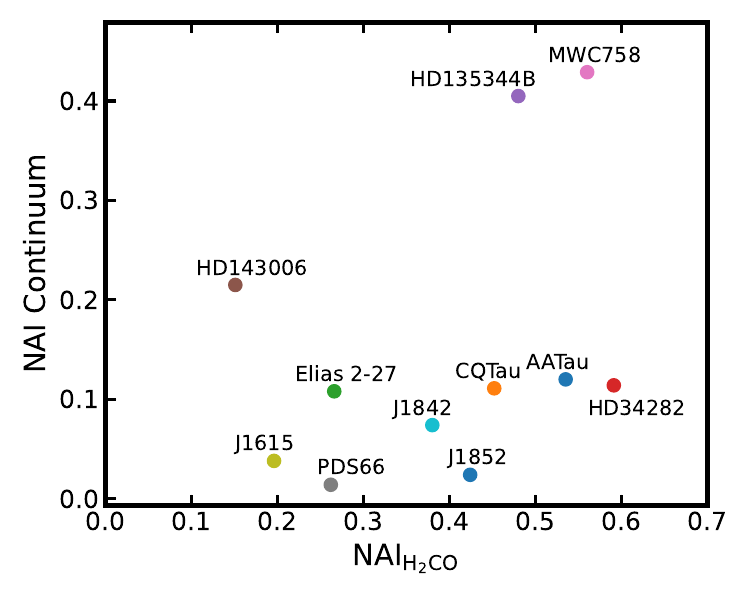}
      \caption{Comparison between the NAI calculated from the 0.8 dust mm continuum emission in exoALMA \citep{exoALMAIV_Curone}, and the one calculated from the peak intensity emission maps of the H$_2$CO 4(0,4)-3(0,3) line. There is no clear correlation between the two indices.
              }
         \label{Fig:NAI}
   \end{figure}


We show a comparison between the NAI values from the continuum data reported in \cite{exoALMAIV_Curone} and the NAI$_{\rm H_2CO}$ values calculated in this work in Figure \ref{Fig:NAI}. The values of dust continuum and NAI$_{\rm H_2CO}$  across the sample do not exhibit a clear correlation. Some disks show high asymmetry in H$_2$CO emission but relatively symmetric dust continuum, while others show the opposite trend. Among the sources with the highest NAI and NAI$_{\rm H_2CO}$ are MWC 758 and HD 135344B, which is not surprising given that both of them show spirals even observed in infrared observations \citep[see, for example, ][]{Ren..2023}. Remarkably, a source with known spiral structure, Elias 2-27, does not have high NAIs in either continuum or line emission based on the metrics presented in this work.  An outstanding disk with shadows and asymmetries in infrared observations \citep{Benisty..2018} and  millimeter dust continuum is HD 143006 \citep{DSHARP..X, exoALMAIV_Curone}; however, it is the source with the lowest NAI$_{\rm H_2CO}$ value and a very smooth formaldehyde emission.

\section{Discussion}\label{Sect: Discussion}

\subsection{Two reservoirs of H$_2$CO}


In our fit, we assume a single-component formaldehyde reservoir, i.e., we fit for only one emitting component of the H$_2$CO lines. However, it is likely that more than one component is present since there is not a unique mechanism to produce formaldehyde, and they are not mutually exclusive. Having more than one component can be explained by a combination of emission arising from warm H$_2$CO being formed by gas-phase reactions, and a colder component produced by ice-surface chemistry and subsequently nonthermally desorbed from the dust grains. The blending of these two components can be linked to the finite spatial resolution of the observation, effectively blending the distinct emitting reservoirs. One possibility is that the colder component is associated with a more radially extended emission, while it could also be a vertical stratification in the disk following a vertical thermal gradient. We highlight that even though we observed four different lines, the lowest upper energy level among them is $\sim$35 K, which makes these lines less sensitive than other studies comprising lower energy lines (typically, the $21\,$K line at 218.222\,GHz) to the outer emission from colder regions close to the CO sublimation temperature. Spatially resolved observations with a larger sample encompassing a broad range of inclinations and stellar spectral types would lead to a better understanding of the COM chemistry in disks.

Recently, \citet{Evans..2025} reported the decrease in the rotational temperature from rotation diagrams fitting between the inner and outer disk of the HD 100456 disk using methanol and formaldehyde lines. Based on the column density ratio between methanol and formaldehyde, they concluded that there are two different emitting reservoirs. In our sample, we cannot conclude that there is a difference in the reservoirs based on excitation temperatures of formaldehyde between the inner ($<$100 au) and outer  (100-200 au) radial bins, due to the large uncertainties. Nevertheless, we see significant differences in the column densities. There is a systematic decrease in the retrieved values from the inner 100 au to the outer region, where the column densities decrease by a factor of almost three. Without specific thermochemical modeling, it is hard to assess whether the trend is dominated by the expected decrease in the gas surface density with increasing radius or by some significant changes in formaldehyde abundance at different radii. Nevertheless, replicating the rotational diagram analysis with multiple tracers observable with optically thin lines can lead to direct estimations of the radial relative abundances or gas surface density.

Previous studies by \cite{Qi..2013} reported that the low excitation temperature of H$_2$CO emission ($\sim$20 K) coincides with a CO ice-regulated chemistry, i.e., formaldehyde is formed through grain-surface reactions by hydrogenation of the CO ice. In fact, we have a large discrepancy with AA Tau, where \cite{Qi..2013} measured $T_{\rm exc}<$20 K, while we retrieved  $T_{\rm exc}>$45 K. We link the discrepancy to a limited sensitivity of the observations from \cite{Qi..2013} compared to the current analyzed dataset, producing larger uncertainties in the retrieved values. Furthermore, our sample expands the explored parameter space in effective temperature and luminosity through more spatially resolved ALMA observations of 11 disks, including four Herbigs. When comparing the retrieved physical conditions of formaldehyde, our findings show a correlation between the effective temperature of the stars and the H$_2$CO excitation temperature (see Fig.  \ref{Fig:corr_all}). Such correlation is not unexpected; even though they are different environments, \cite{Tang..2018} had previously reported that formaldehyde emission typically scales with the bolometric luminosity in galaxies, showing its thermal dependence with the overall radiative energy emission in the system.

Even though the trend shown in Figure \ref{Fig:corr_all} shows that the H$_2$CO excitation temperature increases with the stellar luminosity, it is still hard to fully disentangle the H$_2$CO main formation path. For example, as previously reported by \cite{Oberg..2017}, formaldehyde emission in the TW Hya disk actually traces both distinct reservoirs. For colder stars, the formation of formaldehyde through hydrogenation of CO ices occurs farther out in the disk as the sublimation front is pushed outwards, while for hotter stars, the gas-phase production mechanism extends over a larger region in the inner disk. Nevertheless, we observe that the formaldehyde excitation temperature for multiple disks is quite higher than the CO freeze-out temperature, even beyond 100 au. Only two out of the 11 disks have $T_{\rm exc, out}<25$ K in the outer 100 au. If formaldehyde ice-surface production is significant, these higher excitation temperature conditions require H$_2$CO to be produced on the grains beyond the CO ice line and then transported inwards through pebble drift and being desorbed into the gas phase. Future studies that can spatially resolve the region inside the formaldehyde ice line could provide stronger constraints between the two contributions by isolating the H$_2$CO produced by gas-phase chemistry.

\subsection{Possible Correlations}

Even though we found a possibly promising positive correlation ($\tau>$0)  between the effective temperature and the excitation temperature of formaldehyde, in order to validate the reported relationship, larger and unbiased samples covering a wider range of stellar luminosities and spectral types are required. If the correlation were validated, it would add an additional piece of information to unveil the conditions exciting the emission of COMs' building blocks. The difference in the relationship with respect to previously reported values from unresolved observations showcases ALMA's potential to inform the chemical evolution of disks in different environments.


\subsubsection{NAIs}

Depending on the observables being used, assessing the level of asymmetry in a disk may have different approaches as they can trace different phenomena or range over different values. For example, a natural comparison from this work is the range of values for the NAI in both continuum and gas emission. In general, the range of values of the asymmetry indices in the dust continuum emission is not as wide as that of the ones present in gas emission. We notice that the azimuthal asymmetry index for the gaseous emission has an average value of 0.39, which is significantly higher than the NAI continuum average, which is 0.15. This difference intrinsically points to the fact that  gas emission has more dependencies and degeneracies (chemistry, abundance, dynamics, temperature, emitting layer, among others), while dust emission spans over a narrower range of values with variations that are typically smaller than the ones in line emission. Although there is a lower average for the continuum NAI, the data dispersion has similar values for both samples, with respective standard deviations of 0.15 and 0.14 (see Table \ref{Table: Disks}). However, the sample size needs to be larger for more robust and unbiased constraints. We note that even though both indices range between 0 and 1, the nature of the emission they use and how they are calculated is different, so they should be compared within the same metric with other disks rather than in absolute terms. 

Our measured NAI$_{\mathrm{H_2CO}}$ shows a negative correlation with the flux of the brightest formaldehyde transition within the observed lines (see Figure \ref{Fig:corr_all}). As the H$_2$CO$_{4(0,4)-3(0,3)}$ line flux decreases, any asymmetry in the disk emission translates into a more significant NAI$_{\mathrm{H_2CO}}$ since the index is normalized by the total flux. If the line emission is weak across the disk, localized spots of strong emission will have a more significant imprint that will be retrieved by our proposed metric. On the contrary, slight asymmetries in disks with strong line emission are weakened by the normalization factor, as expected.

Introducing a formal index definition provides a grounding point for comparison. Figure \ref{Fig:MOM0} shows the peak emission of the brightest observed H$_2$CO line. Visually, PDS 66 and Elias 2-27 may seem quite asymmetric, although in absolute values and compared to the sensitivity of the observations, such asymmetries may not translate to a particularly high NAI$_{\mathrm{H_2CO}}$. Other disks with faint emission, such as AA Tau or J1852, show high levels of asymmetry within the sample. Thus, the metric informs about an intrinsic behavior that cannot be directly inferred just by looking at the emission by itself, as it considers relative emission compared to the line flux and also takes into account the sensitivity level of the observations by not confusing the noise random distribution as a possible source of asymmetry. Among the disks with large spirals, Elias 2-27 is quite symmetric compared to MWC 758 and HD 135344B. Even when its continuum emission is visually asymmetric, the deviations are at a very low percentage level compared to the average value. The main difference between Elias 2-27, MWC 758 and HD 135344B is their age, with Elias 2-27 being the youngest among them. It is possible that the H$_2$CO asymmetries may develop over longer timescales. Another possibility is that the disks with large spirals in the sample, except for Elias 2-27, have deep cavities in the continuum, which can enhance the value of the measures NAI in the continuum. In any case, a larger and more meaningful sample is required to verify whether or not these are valid trends.

As the spectral bandwidth is being increased in the near future, studies such as this, obtaining multiple transitions of the same molecule in one scheduling block, will enable the community to retrieve the physical parameters across different molecular tracers. THus, we will be able to link those parameters to other disk and stellar properties, providing key information about the chemical evolution of the gas reservoir in planet-forming environments. Rotational diagram analysis with good sensitivity at high spatial resolution will enable direct characterization of the different components traced by different COMs, and also the OPR of hydrogen using formaldehyde, for example.

\subsection{Effect of the OPR}

In our analysis, we assumed an OPR=3. However, observations in molecular clouds have shown that it can reach lower values through hydrogen fractionation in earlier stages. TW Hya, for example, has a nominal OPR $\approx$3 in the inner regions and starts decreasing to values between 1 and 2 at 180 au \citep{van..Scheltinga..21}. Assuming a lower OPR would make the slope of the fit in the rotational diagrams steeper in the transitions with higher upper energy. Thus, the retrieved excitation temperature would be lower. Even though changes in the assumed OPR would affect the exact retrieved values, the trends would be similar to those inferred in our analysis. However, we cannot disregard the fact that possible variations in H$_2$CO along different spectral types and/or environmental conditions cannot lead to different trends compared to the ones found in this sample. Given that the two para-H$_2$CO lines have very similar quantum numbers and upper state energy levels, the amount of information on the OPR that can be retrieved from them is limited. Even though a different OPR would change the excitation temperature, it would also provide interesting insights regarding the chemical evolution of hydrogen in the different regions. A radially variable OPR can have significant consequences for the chemistry at different disk locations. In such cases, the physical conditions of the gas with different nuclear spins should be determined independently. Thus, providing a larger statistical sample with multiple ortho- and para-formaldehyde transitions across multiple regions will lead to a better understanding of COMs chemistry in different formation environments.

\section{Summary}\label{Sect: Summary}
 
   In this work, we assessed possible correlations between resolved formaldehyde emission and disk/stellar properties, attempting to understand the origin of the formaldehyde reservoirs and their production mechanisms. We also explored relationships between the axisymmetry from high-resolution dust continuum observations within the exoALMA sample as a proxy of dynamical state, and compared it with the asymmetry observed with H$_2$CO line emission. We aimed to determine if there is a clear correlation that can link the dust structure and dynamical states with the chemical evolution in the sample. Our conclusions are as follows:
   \begin{enumerate}
      \item We do not find sufficient evidence to establish statistically significant correlations. Nevertheless, we find suggestive evidence indicative of a possible correlation between the stellar effective temperature and the excitation temperature of formaldehyde within the inner 100 au, potentially linking COM chemical pathways to stellar spectral type.
      \item Within the sample, most disks show a decreasing trend for the column density from the innermost 100 au radial bin toward larger radii. Regarding the excitation temperature, due to the uncertainties in our measurements, we cannot draw statistically significant conclusions about variations between the inner 100 au and the regions farther out.
      \item No correlation is found between the NAI from millimeter dust continuum images and the NAI$_{\mathrm{H_2CO}}$ from formaldehyde  emission maps. This suggests that, within the analyzed subsample, there is no clear link between the dynamical state of the disks and their chemical evolution as traced by spatially resolved H$_2$CO emission.

   \end{enumerate}

\newpage

\begin{acknowledgements}

FA, SF and LR are funded by the European Union (ERC, UNVEIL, 101076613). Views and opinions expressed are however those of the author(s) only and do not necessarily reflect those of the European Union or the European Research Council. Neither the European Union nor the granting authority can be held responsible for them. SF acknowledges financial contribution from PRIN-MUR 2022YP5ACE. PC acknowledges support by the Italian Ministero dell'Istruzione, Università e Ricerca through the grant Progetti Premiali 2012 – iALMA (CUP C52I13000140001) and by the ANID BASAL project FB210003. JB acknowledges support from NASA XRP grant No. 80NSSC23K1312. MB and JS have received funding from the European Research Council (ERC) under the European Union’s Horizon 2020 research and innovation programme (PROTOPLANETS, grant agreement No. 101002188). CH gratefully acknowledges support from the U.S. National Science Foundation Grants 2511673 and 2407679, NRAO SOSPADA-036, National Geographic Society, and the Georgia Museum of Natural History. GL has received funding from the European Union's Horizon 2020 research and innovation program under the Marie Sklodowska-Curie grant agreement No. 823823 (DUSTBUSTERS). C.P. acknowledges Australian Research Council funding  via FT170100040, DP18010423, DP220103767, and DP240103290. JDI acknowledges support from an STFC Ernest Rutherford Fellowship (ST/W004119/1) and a University Academic Fellowship from the University of Leeds.
This paper makes use of the following ALMA data:\\ADS/JAO.ALMA\#2016.1.00484.L,\\ ADS/JAO.ALMA\#2021.1.01123.L,\\ ADS/JAO.ALMA\#2022.1.00485.S\\ and ADS/JAO.ALMA\#2023.1.00334.S.\\ ALMA is a partnership of ESO (representing its member states), NSF (USA) and NINS (Japan), together with NRC(Canada), MOST and ASIAA (Taiwan), and KASI (Republic of Korea), in cooperation with the Republic of Chile. The Joint ALMA Observatory is operated by ESO, AUI/NRAO and NAOJ. The National Radio Astronomy Observatory is a facility of the National Science Foundation operated under cooperative agreement by Associated Universities, Inc. 

\end{acknowledgements}
%


\software{\texttt{CASA} \citep{CASA2}
\texttt{NumPy} \citep{numpy},
\texttt{matplotlib} \citep{matplotlib},
\texttt{cmasher} \citep{cmasher},
\texttt{scikit-learn} \citep{scikit-learn},
\texttt{statsmodels} \citep{statsmodels},
\texttt{frank} \citep{Jennings..2020}.
          }



\appendix

\section{NAI of Elias 2-27}\label{App: NAI}

In addition to the sources from the exoALMA sample, we included Elias 2-27 and applied the same procedure as outlined by \cite{exoALMAIV_Curone} to retrieve the NAI from the continuum observations. The analyzed data are from the DSHARP Large Program \citep{DSHARP,DSHARP..X}. The residual SNR map and the radial profile extracted with the \texttt{frank} fit \citep{Jennings..2020} and the CLEAN algorithm \citep{Hogbom..74} are shown in Figure \ref{fig:Elias_NAI}. Despite the noticeable spiral structure in the residual SNR map in Fig. \ref{fig:Elias_NAI}, the amplitude of the deviations with respect to the azimuthally-averaged profile is relatively small. Thus, its NAI has a lower value than the average of the exoALMA sources.

\begin{figure}
    \centering
    \includegraphics[height=0.33\textheight]{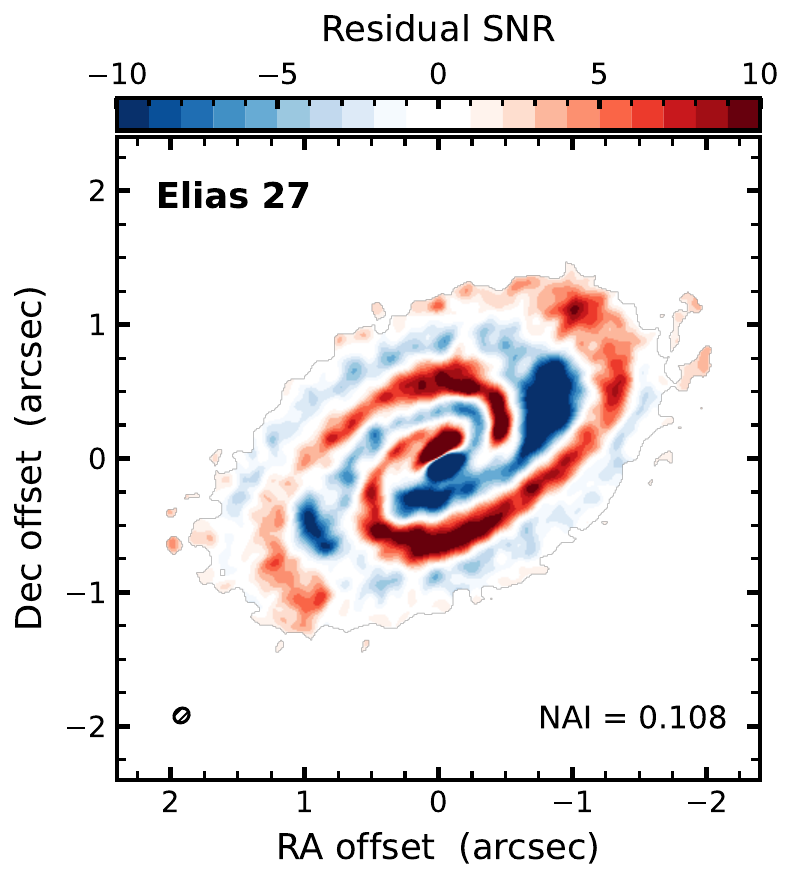}
    \includegraphics[height=0.29\textheight]{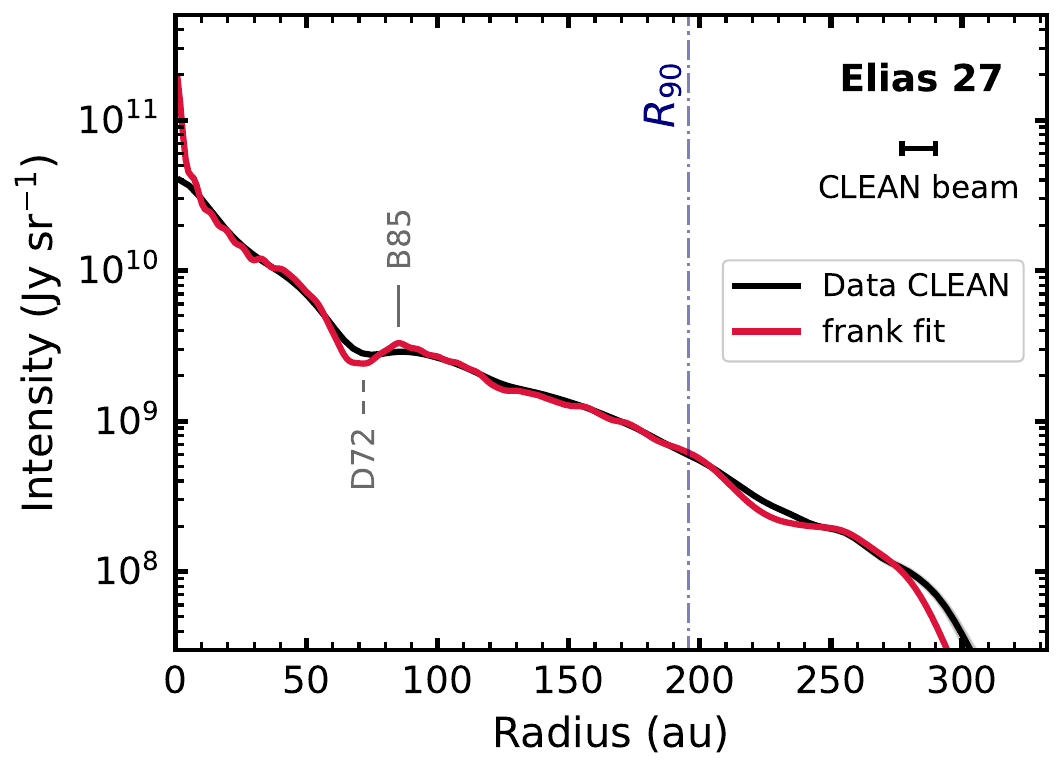}
    \caption{\textbf{Left:} Residuals obtained subtracting an axisymmetric \texttt{frank} model from the continuum data emission of Elias 2-27. \textbf{Right:} Radial profile from the \texttt{frank} model and with the CLEAN algorithm. Non-axisymmetry index is calculated following the same procedure detailed in \citet{exoALMAIV_Curone}.}
    \label{fig:Elias_NAI}
\end{figure}

\section{Integrated and Peak Intensity Maps}\label{App: Maps}

We show the additional integrated intensity (Moment 0) and peak intensity (Moment 8) maps of all the lines in Figures \ref{Fig:MOM_2} and \ref{Fig:MOM_3}. The integrated intensity  maps were used to retrieve the physical properties of the formaldehyde emission from the MCMC fitting of the rotation diagrams, while the peak intensity maps were used to calculate the level of asymmetry in the gas emission.

\begin{figure*}
   \centering
   \includegraphics[width=0.495\textwidth]{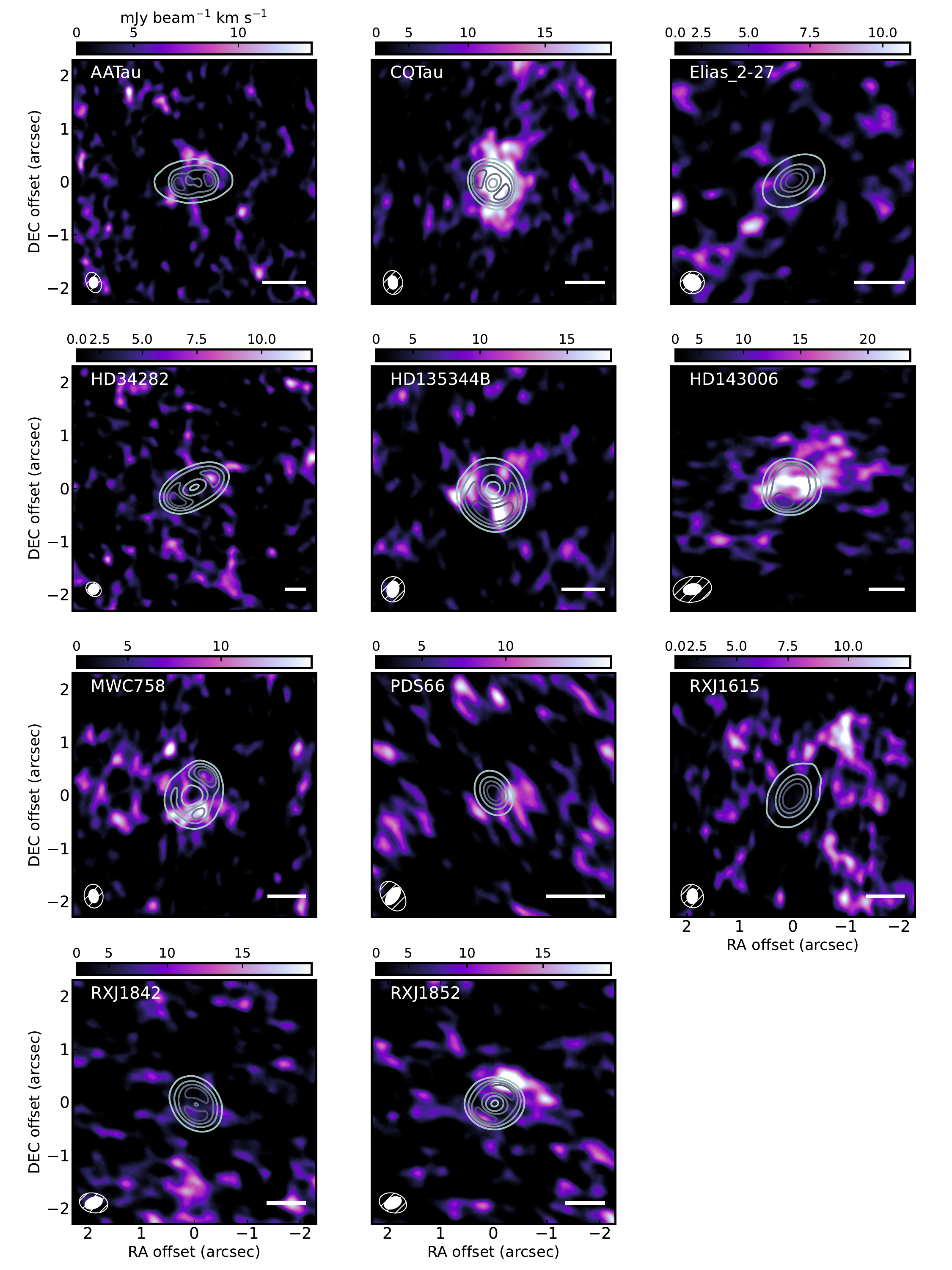}
   \includegraphics[width=0.495\textwidth]{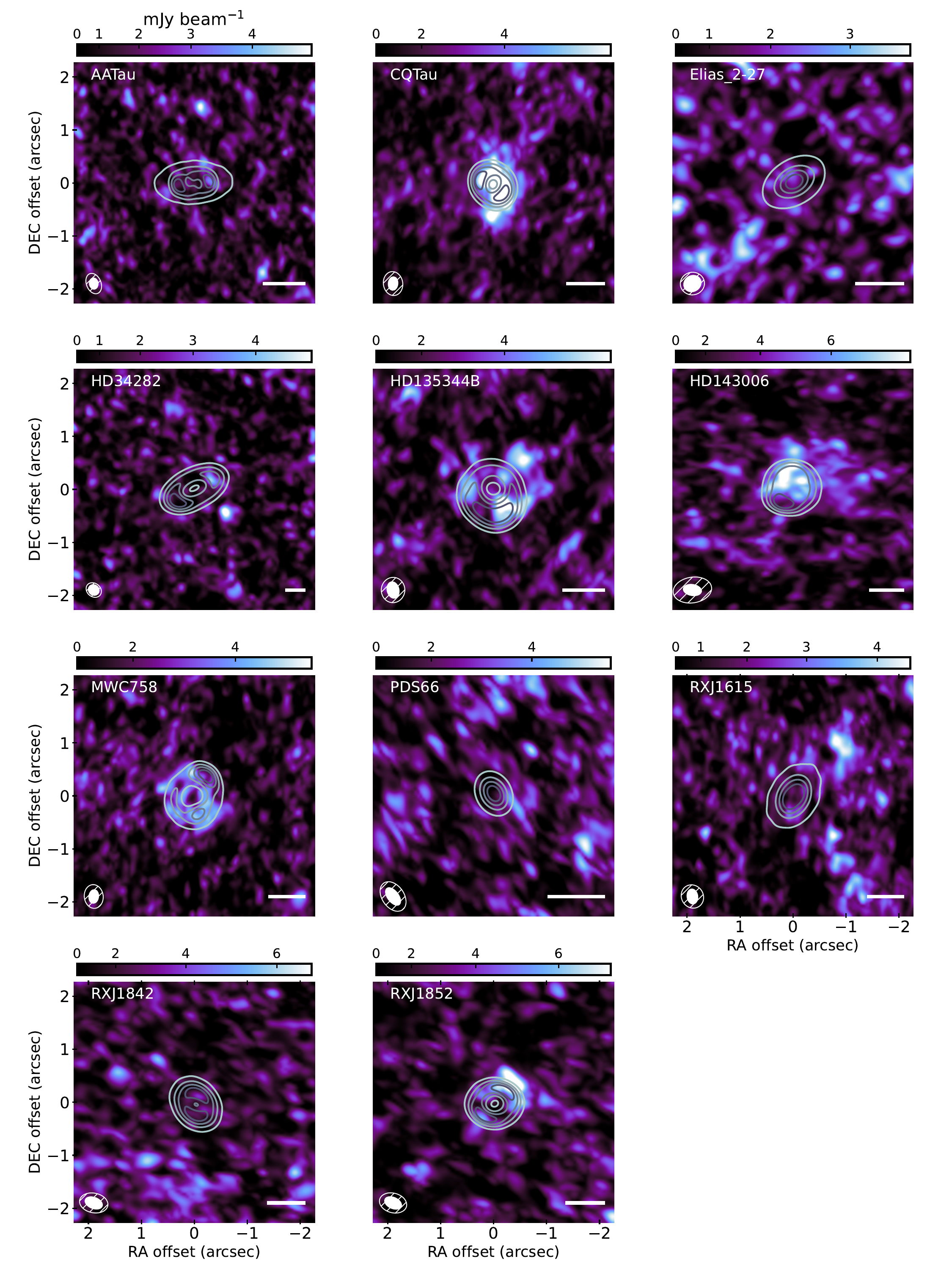}
      \caption{Same as Figure \ref{Fig:MOM0} but for the H$_2$CO 4(2,3)-3(2,2) line.
              }
         \label{Fig:MOM_2}
   \end{figure*}

\begin{figure*}
   \centering
   \includegraphics[width=0.495\textwidth]{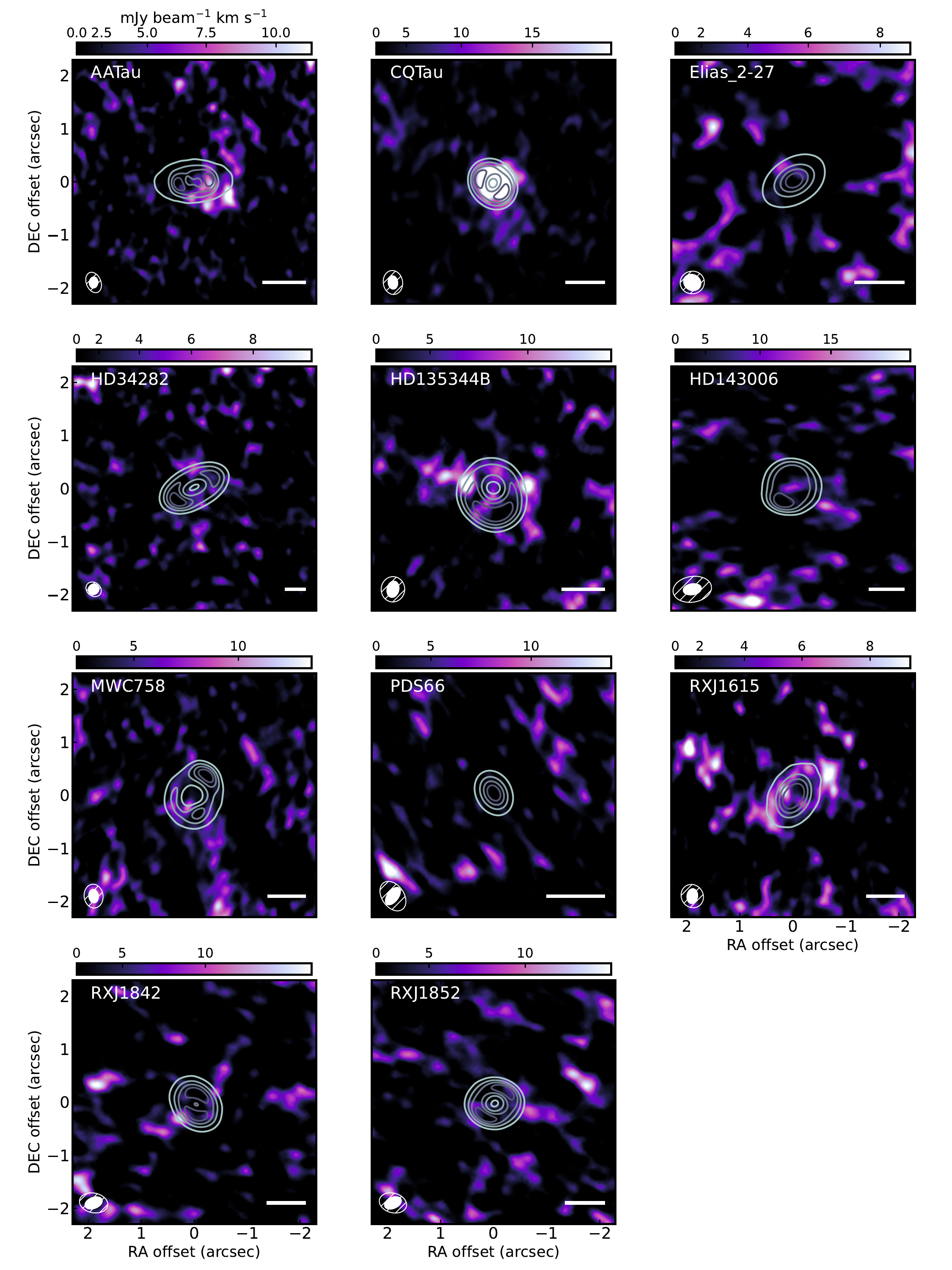}
   \includegraphics[width=0.495\textwidth]{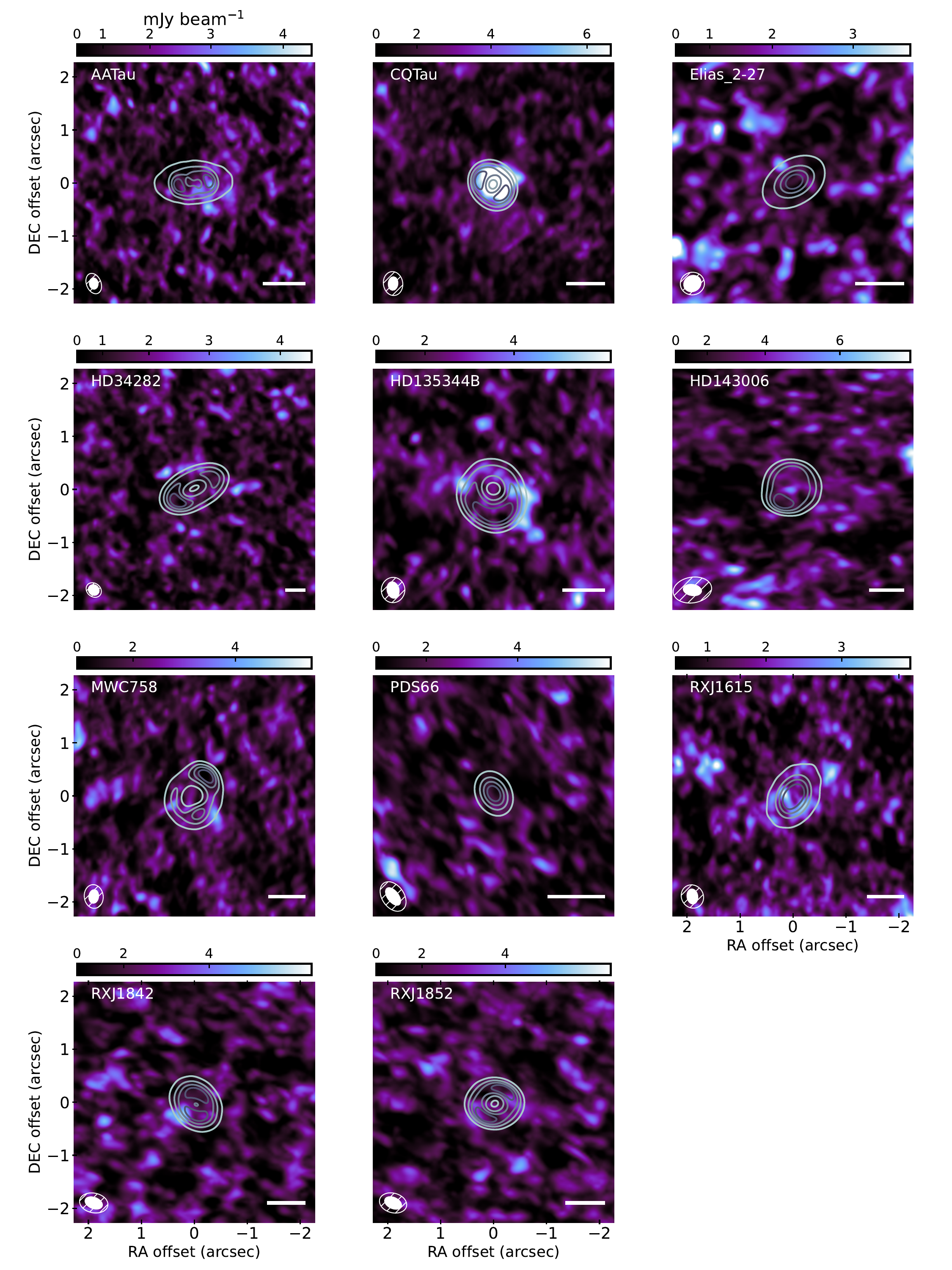}
      \caption{Same as Figure \ref{Fig:MOM0} but for the H$_2$CO 4(3,2)-3(3,1) and 4(3,1)-3(3,0) lines. In this case we show the average of the two respective moment maps of each line.}
         \label{Fig:MOM_3}
   \end{figure*}

\begingroup
\renewcommand{\arraystretch}{1.3}
 \begin{table*}
      \caption{Targeted disks and retrieved physical parameters of formaldehyde emission in annuli between 100 and 200 au.}
         \label{Table: Disks_out}
     $$ 
         \begin{array}{lccccc}
            \hline
            \noalign{\smallskip}
             \mathrm{Disk} &  F_{\mathrm{H_2CO}, 4_{0,4}-3_{0,3}}  & F_{\mathrm{H_2CO}, 4_{2,3}-3_{2,2}}  & F_{\mathrm{H_2CO},  4_{3,2}-3_{3,1}, 4_{3,1}-3_{3,0}}  &   \mathrm{T_{exc, out}} & \mathrm{log(N_{H_2CO, out})}\\
            &  \mathrm{(mJy\ km\ s^{-1})} & \mathrm{(mJy\ km\ s^{-1})}   & \mathrm{(mJy\ km\ s^{-1})}  & \mathrm{(K)} & \mathrm{(cm^{-2})} \\
            \noalign{\smallskip}
            \hline
            \noalign{\smallskip}
            \mathrm{AA \ Tau} & 90.0 \pm 15.0 & 7.7 \pm 12.2 & 25.4 \pm 12.0 & 49.4^{+30.3}_{-25.9} & 12.75^{+0.16}_{-0.13} \\
            \mathrm{CQ\ Tau} &  148.6 \pm 17.7 & 90.0 \pm 16.1 & 44.9 \pm 18.4 & 91.0^{+23.4}_{-19.3} & 13.25^{+0.10}_{-0.10}\\
            \mathrm{HD\ 34282} &   60.3 \pm 7.8 &  12.0 \pm 9.0 & 12.0 \pm 5.6 & 52.6^{+17.1}_{-15.8} & 13.30^{+0.10}_{-0.10} \\
            \mathrm{HD\ 135344B} & 139.6 \pm 33.6  &  <19.2 & 20.1 \pm 19.9 & 25.5^{+27.6}_{-15.2}  & 12.57^{+0.22}_{-0.16} \\
            \mathrm{HD\ 143006} &  256.6 \pm 18.1 & 44.3 \pm 20.8 &  <12.9 & 26.9^{+7.1}_{-8.5}  & 13.01^{+0.05}_{-0.04} \\
            \mathrm{MWC\ 758} &  88.0 \pm 23.4 & 23.3 \pm 20.4 & 19.4 \pm 18.1 & 65.2^{+40.4}_{-31.2}  & 12.65^{+0.18}_{-0.19} \\
            \mathrm{PDS\ 66} &  62.8 \pm 22.9 & 10.4 \pm 21.9 & 14.3 \pm 21.8 & 65.1^{+47.2}_{-40.7}  & 12.10^{+0.26}_{-0.32} \\
            \mathrm{RX\ J1615} &  217.7 \pm 17.9 & 24.8 \pm 23.9 & 19.1 \pm 9.9 & 33.6^{+8.4}_{-11.6}  & 13.05^{+0.06}_{-0.05} \\
            \mathrm{RX\ J1842} &  155.8 \pm 21.9 & 6.0 \pm 14.6 & <19.3 & 17.8^{+10.9}_{-9.3}  & 12.82^{+0.34}_{-0.09}\\
            \mathrm{RX\ J1852} &  155.8 \pm 6.0 & 47.6 \pm 11.7 & <19.3  & 39.7^{+15.0}_{-14.3} & 12.85^{+0.10}_{-0.08}\\
            \mathrm{Elias\ 2-27} & 255.6 \pm 9.7 & 12.5 \pm 6.0 & <6.2 & 16.7^{+3.1}_{-3.9}  &12.93^{+0.08}_{-0.03} \\

            \noalign{\smallskip}
            \hline
         \end{array}
     $$ 
     \tablecomments{Flux uncertainties do not include the uncertainty in the absolute flux calibration of the ALMA interferometer. The fourth column is the average flux between the H$_2$CO $4_{3,2}-3_{3,1} \mathrm{and}  4_{3,1}-3_{3,0}$ lines.}
\end{table*}

\section{Rotation Diagrams}\label{Appendix: rot}

From Figure \ref{Fig:corner_aatau} to Figure \ref{Fig:corner_elias} we show the corner plots of the MCMC fits for all the disks in the sample. 

\begin{figure*}
   \centering
   \includegraphics[width=0.48\textwidth]{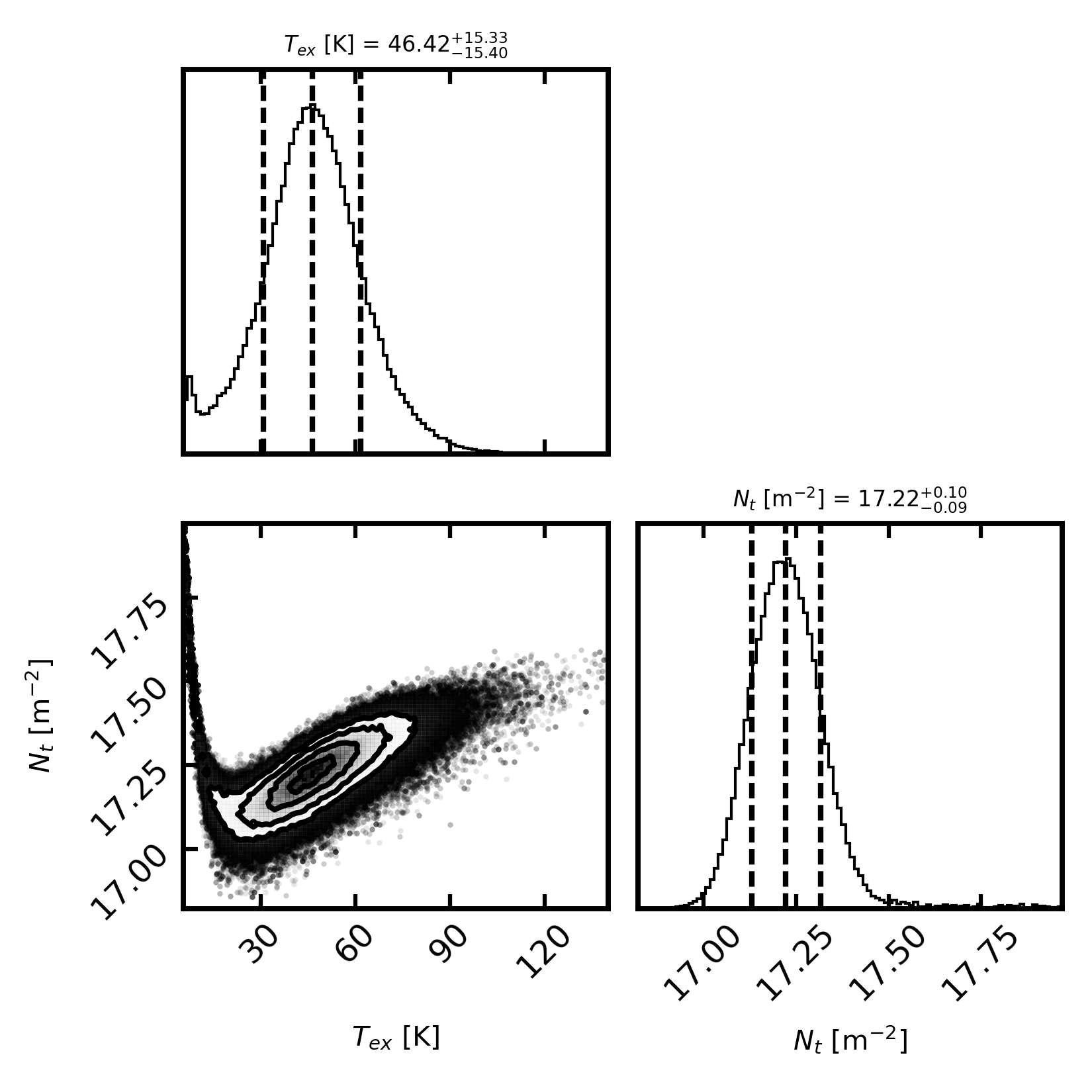}
   \includegraphics[width=0.48\textwidth]{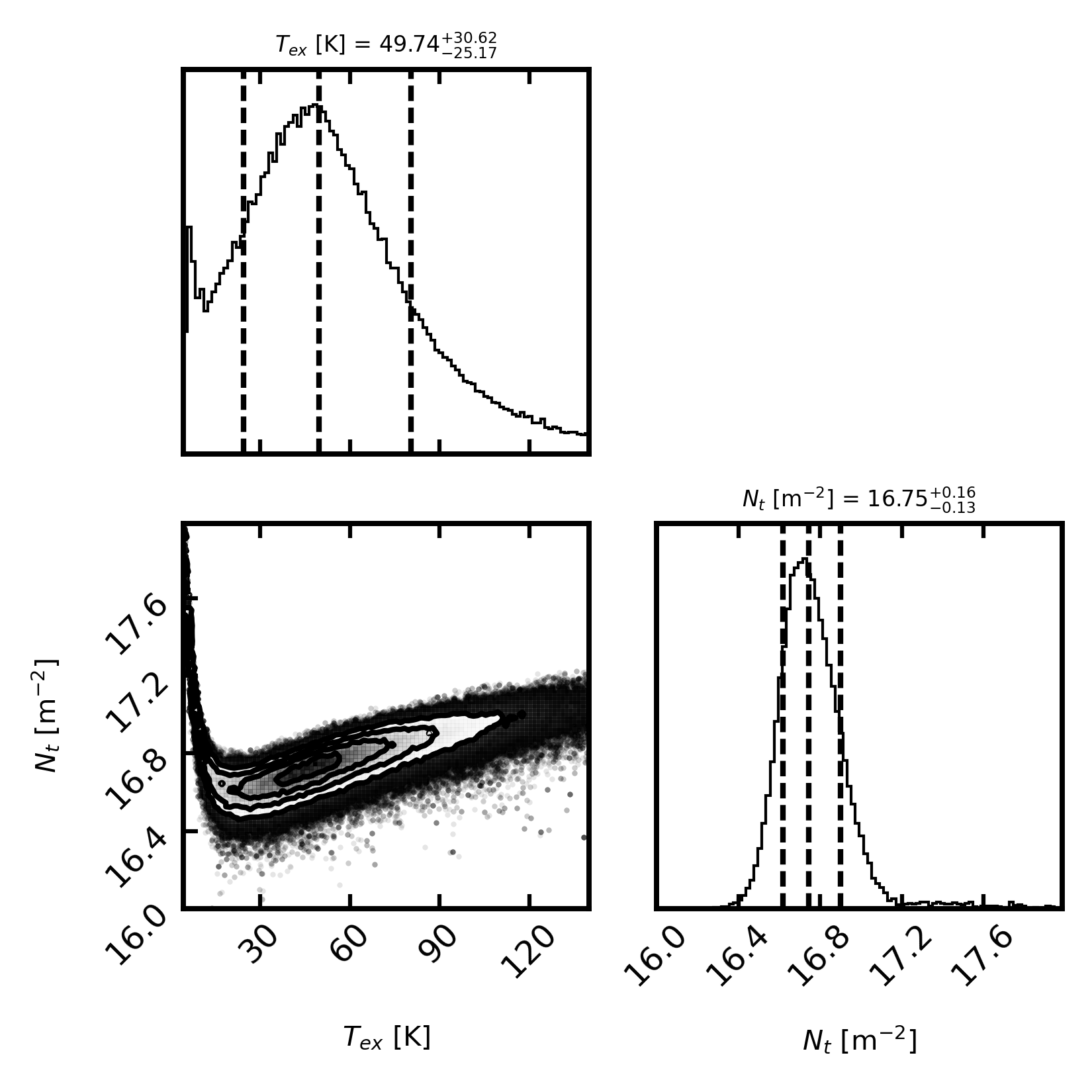}
      \caption{Corner plots of the MCMC fitting of the rotational diagrams of AA Tau. \textbf{Left}: Inner 100 au. \textbf{Right}: 100-200 au annular region.
              }
         \label{Fig:corner_aatau}
\end{figure*}

\begin{figure*}
   \centering
   \includegraphics[width=0.48\textwidth]{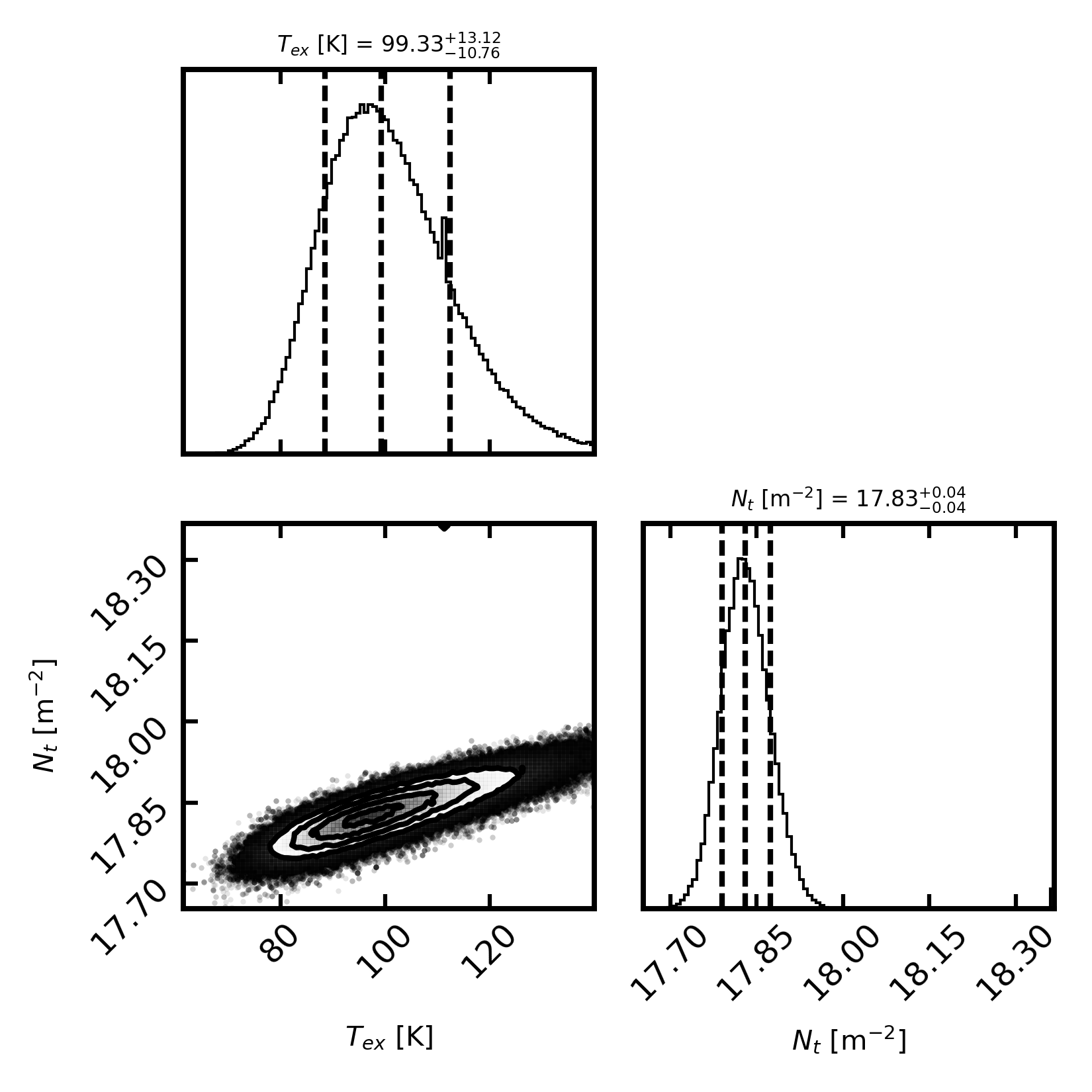}
   \includegraphics[width=0.48\textwidth]{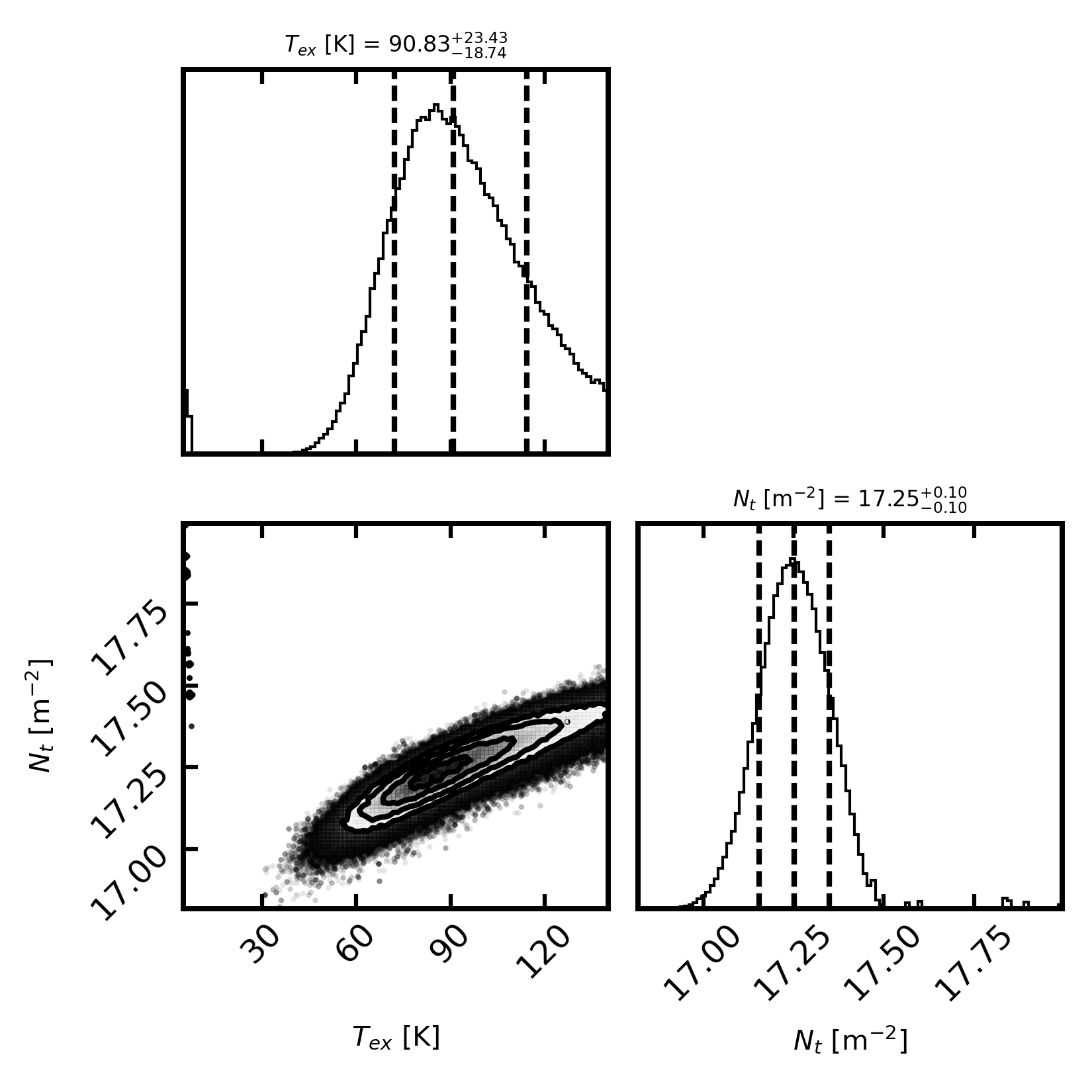}
      \caption{Same as Figure \ref{Fig:corner_aatau} for CQ Tau.
              }
         \label{Fig:corner_cqtau}
\end{figure*}

\begin{figure*}
   \centering
   \includegraphics[width=0.48\textwidth]{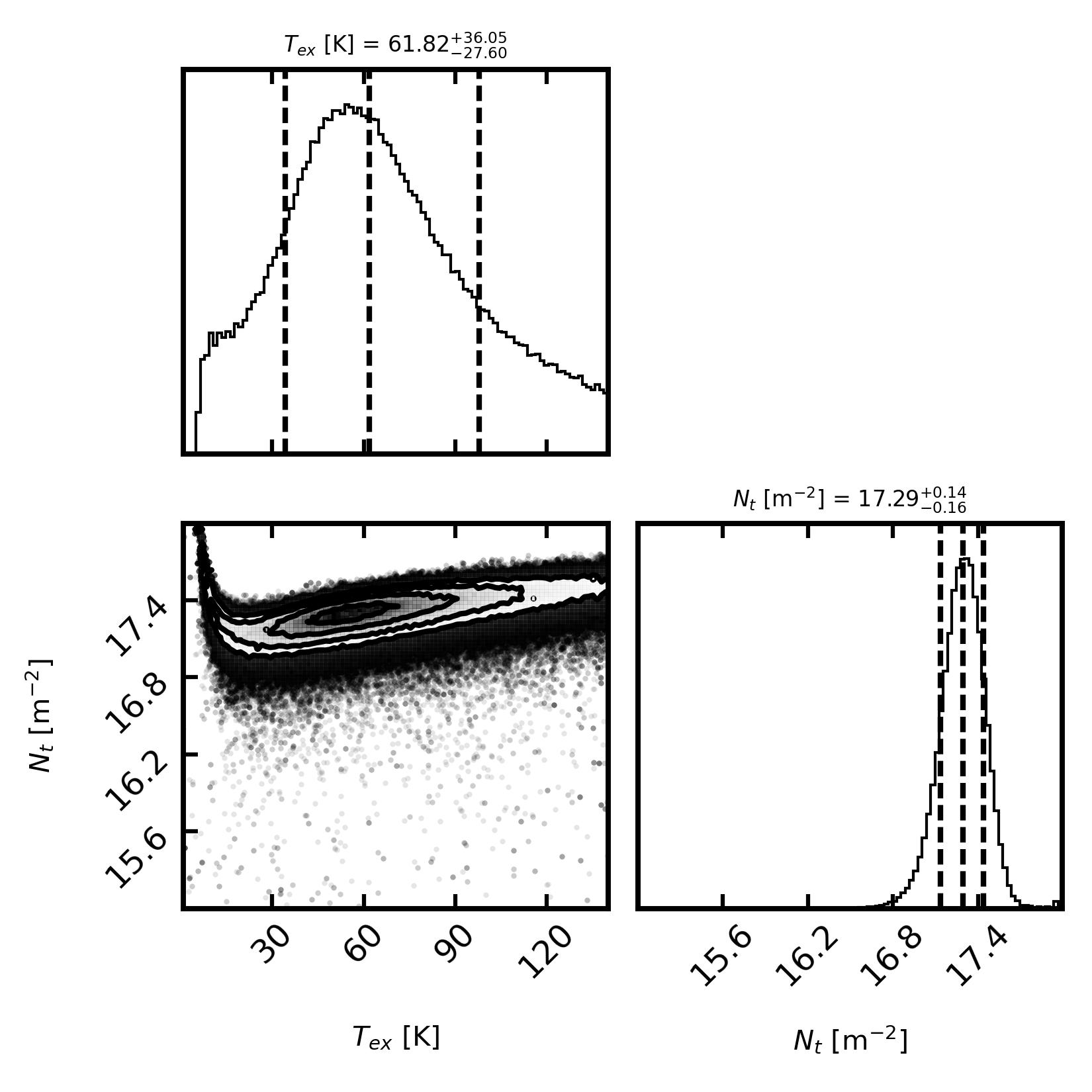}
   \includegraphics[width=0.48\textwidth]{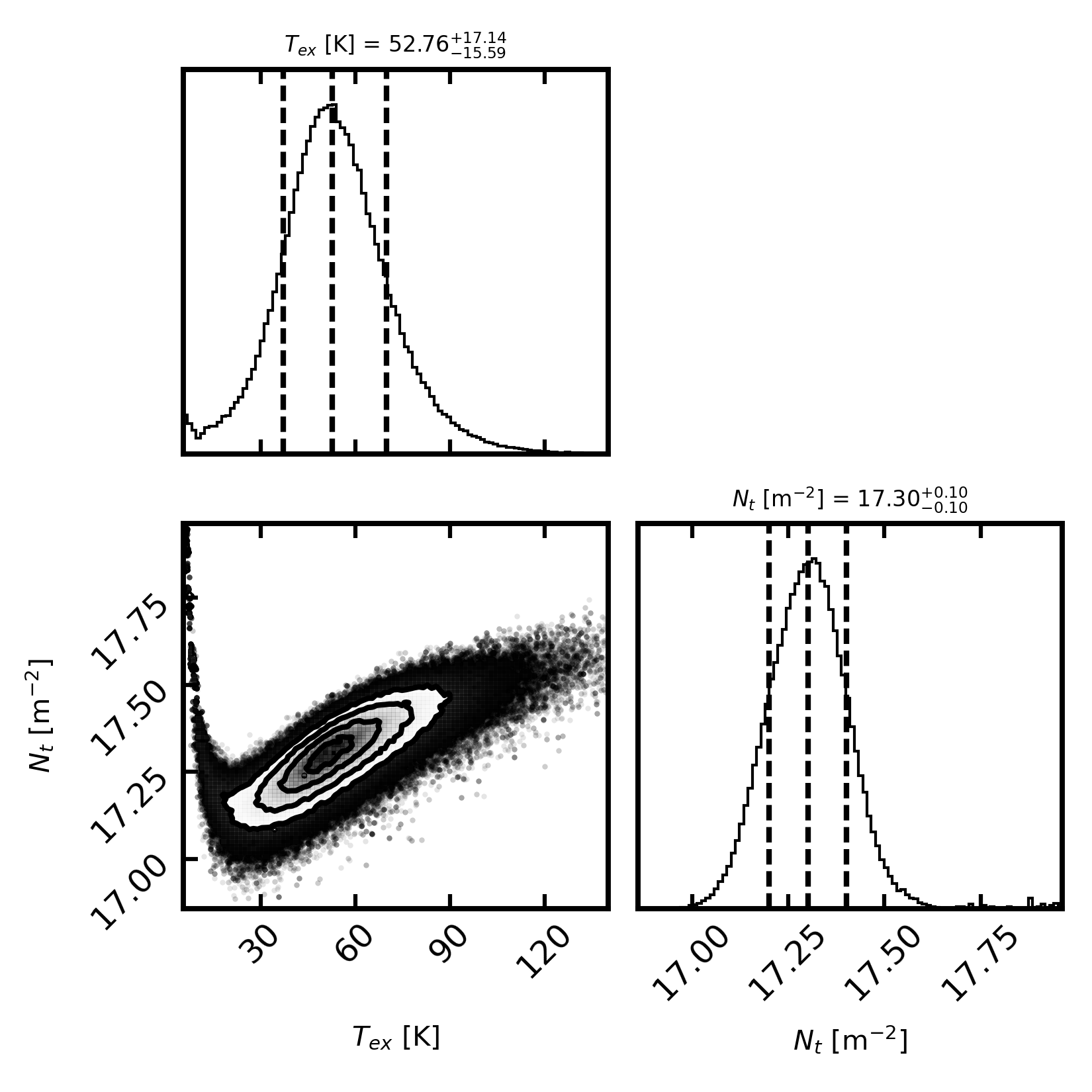}
      \caption{Same as Figure \ref{Fig:corner_aatau} for HD 34282.
              }
         \label{Fig:corner_HD34282}
\end{figure*}

\begin{figure*}
   \centering
   \includegraphics[width=0.48\textwidth]{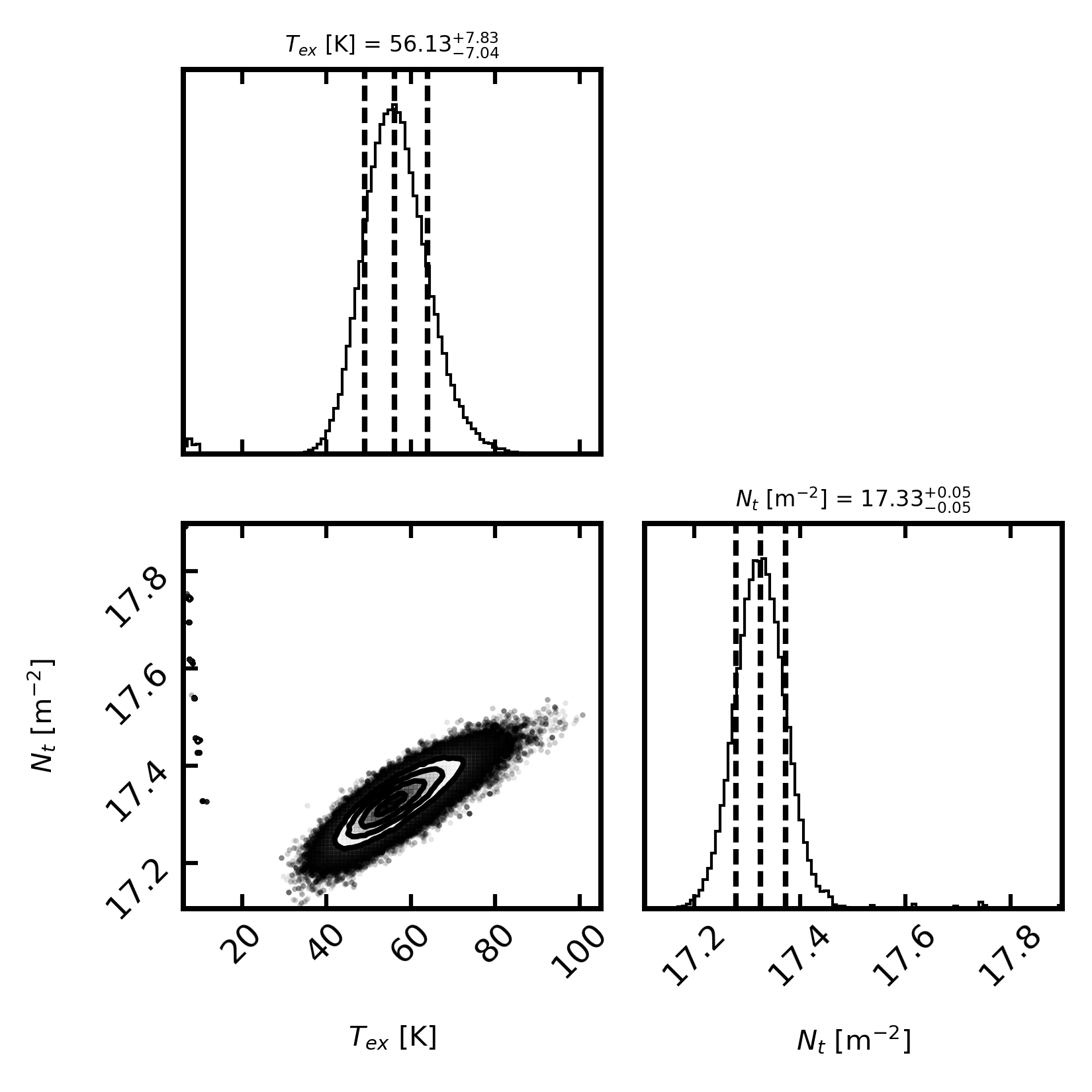}
   \includegraphics[width=0.48\textwidth]{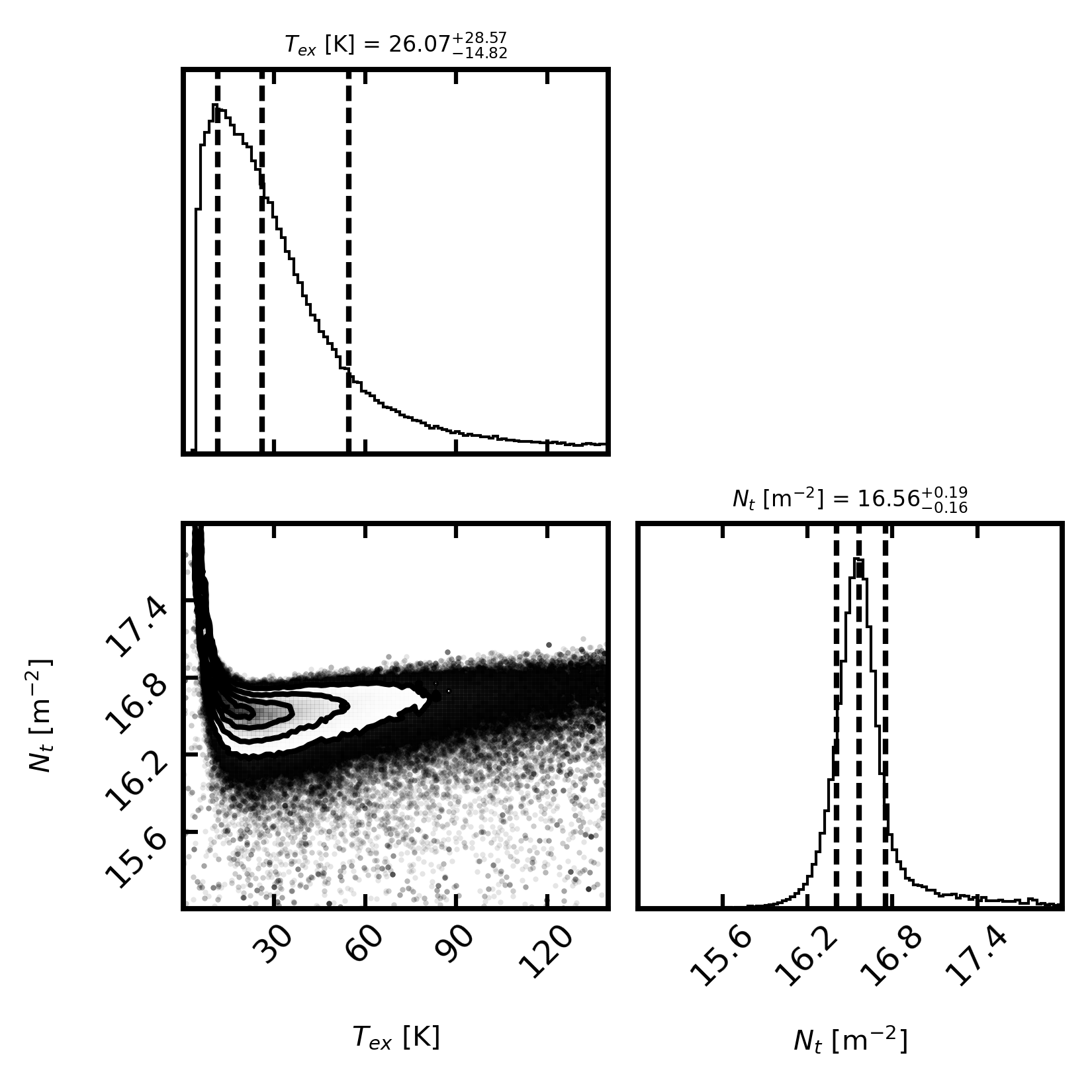}
      \caption{Same as Figure \ref{Fig:corner_aatau} for HD 135344B.
              }
         \label{Fig:corner_HD135344B}
\end{figure*}

\begin{figure*}
   \centering
   \includegraphics[width=0.48\textwidth]{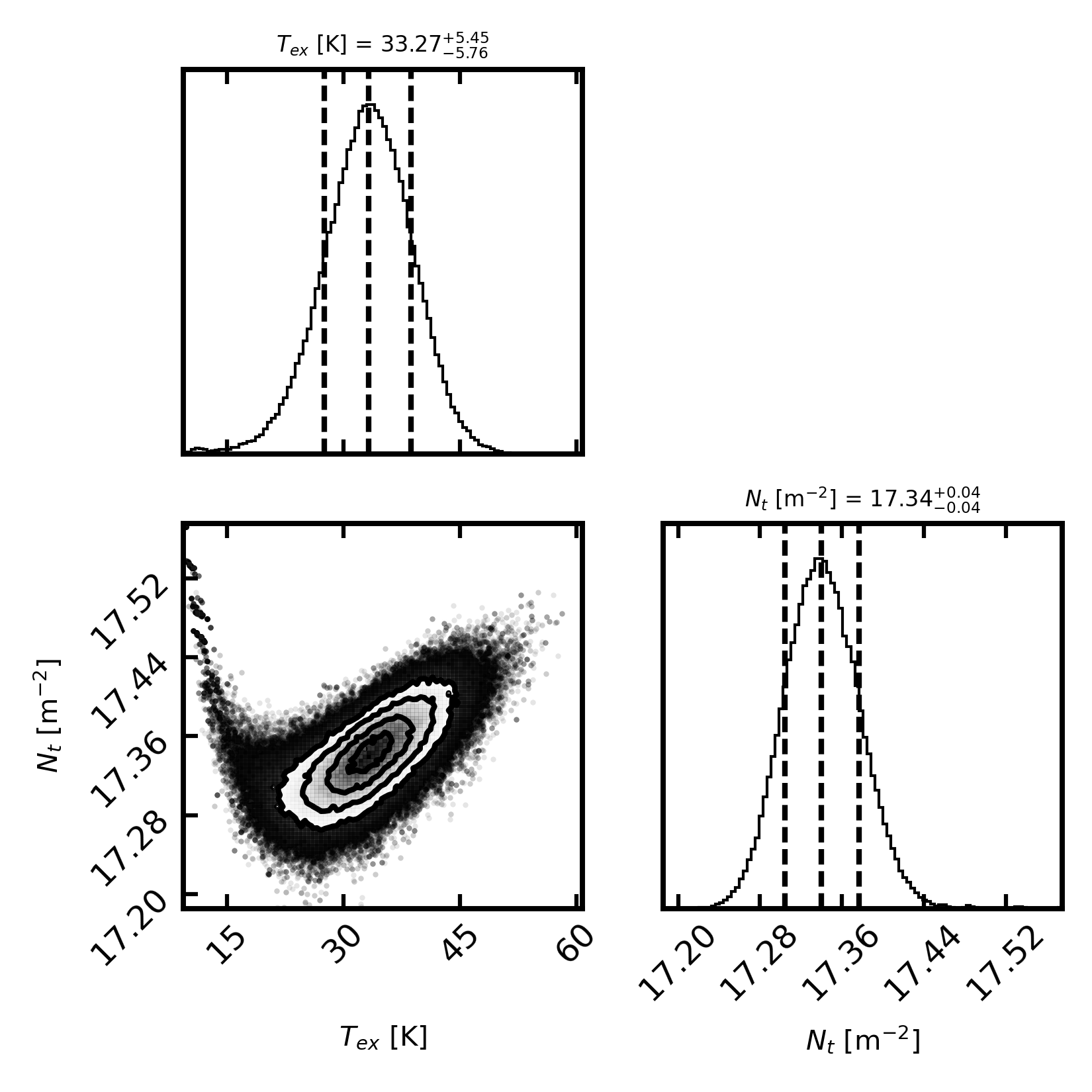}
   \includegraphics[width=0.48\textwidth]{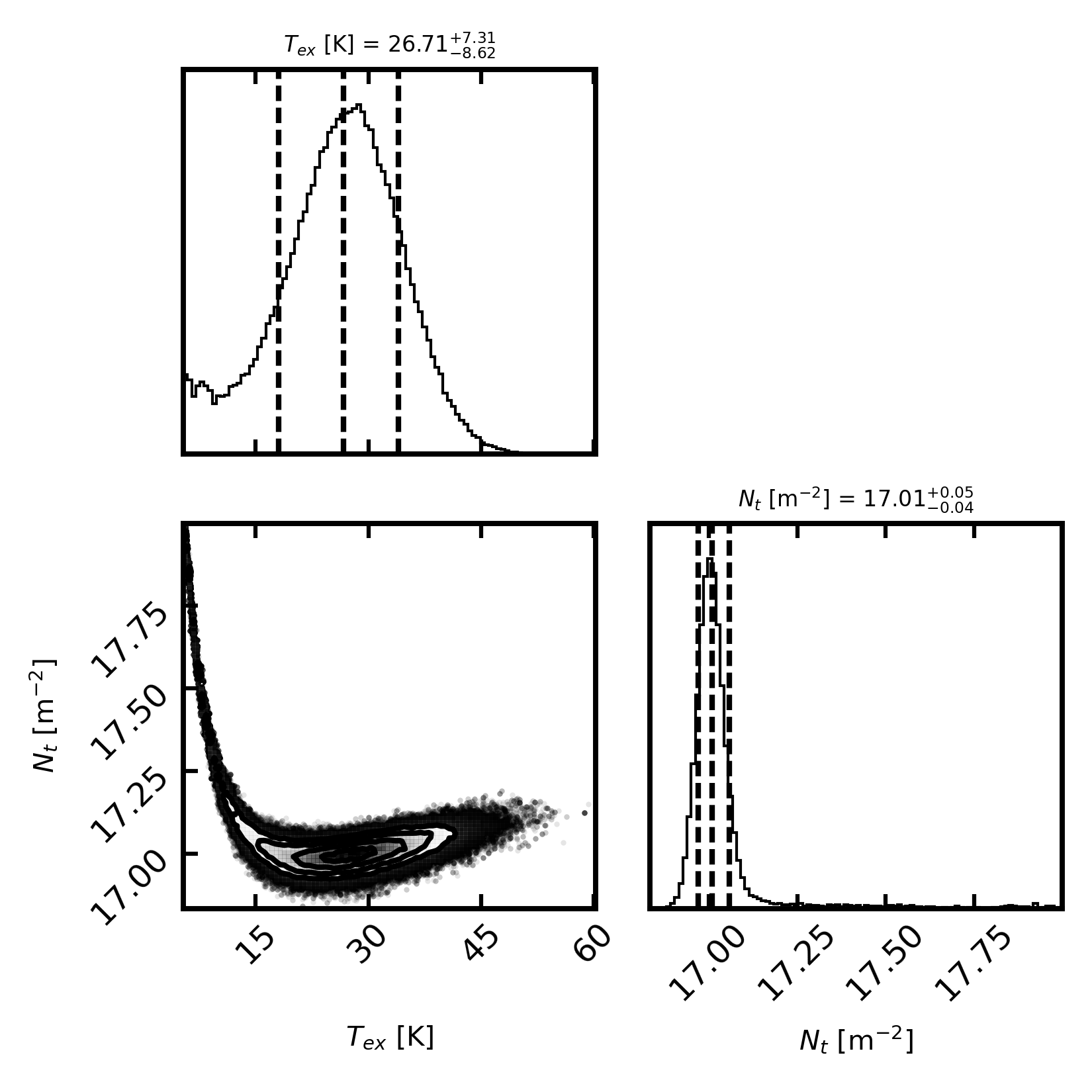}
      \caption{Same as Figure \ref{Fig:corner_aatau} for HD 143006.
              }
         \label{Fig:corner_HD143006}
\end{figure*}

\begin{figure*}
   \centering
   \includegraphics[width=0.48\textwidth]{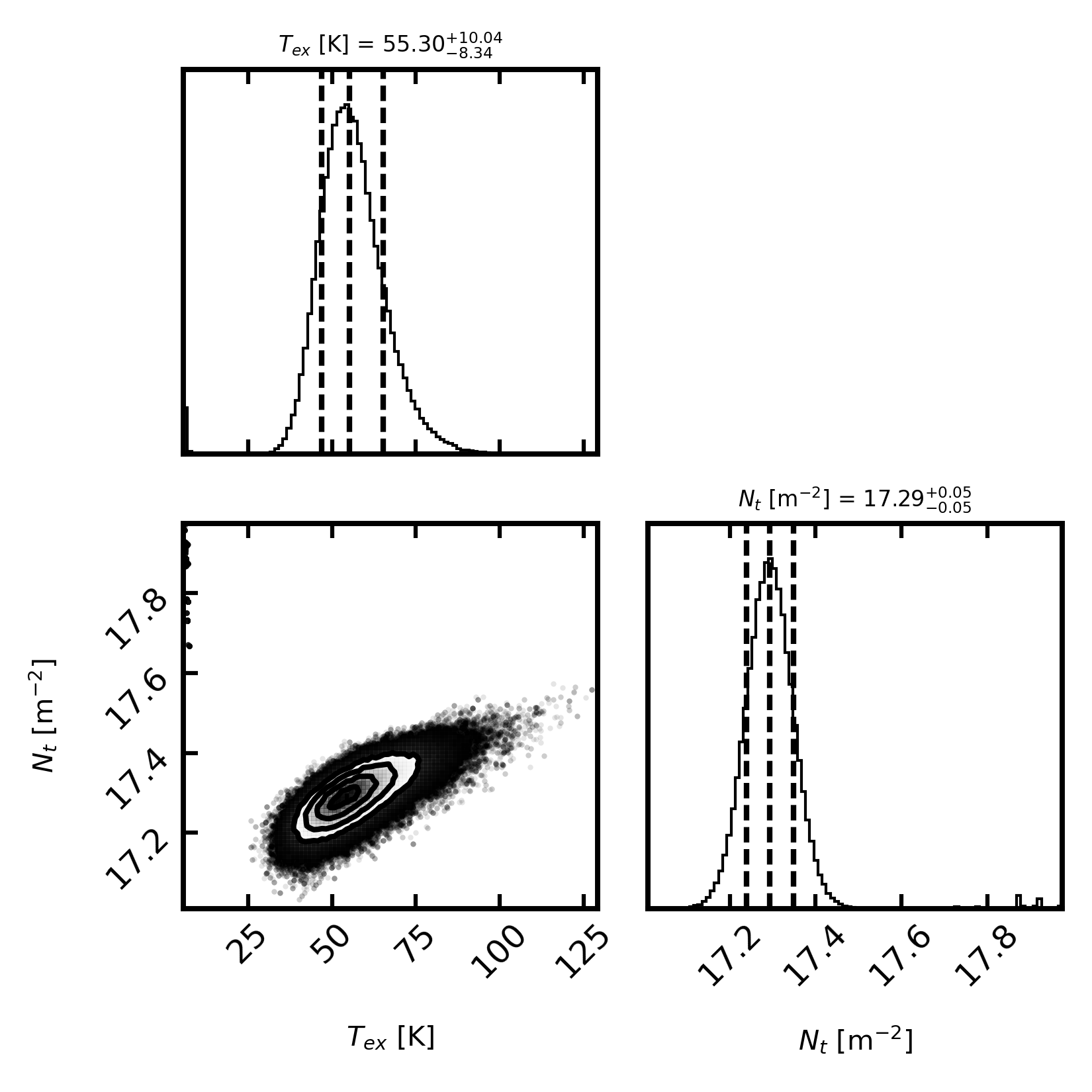}
   \includegraphics[width=0.48\textwidth]{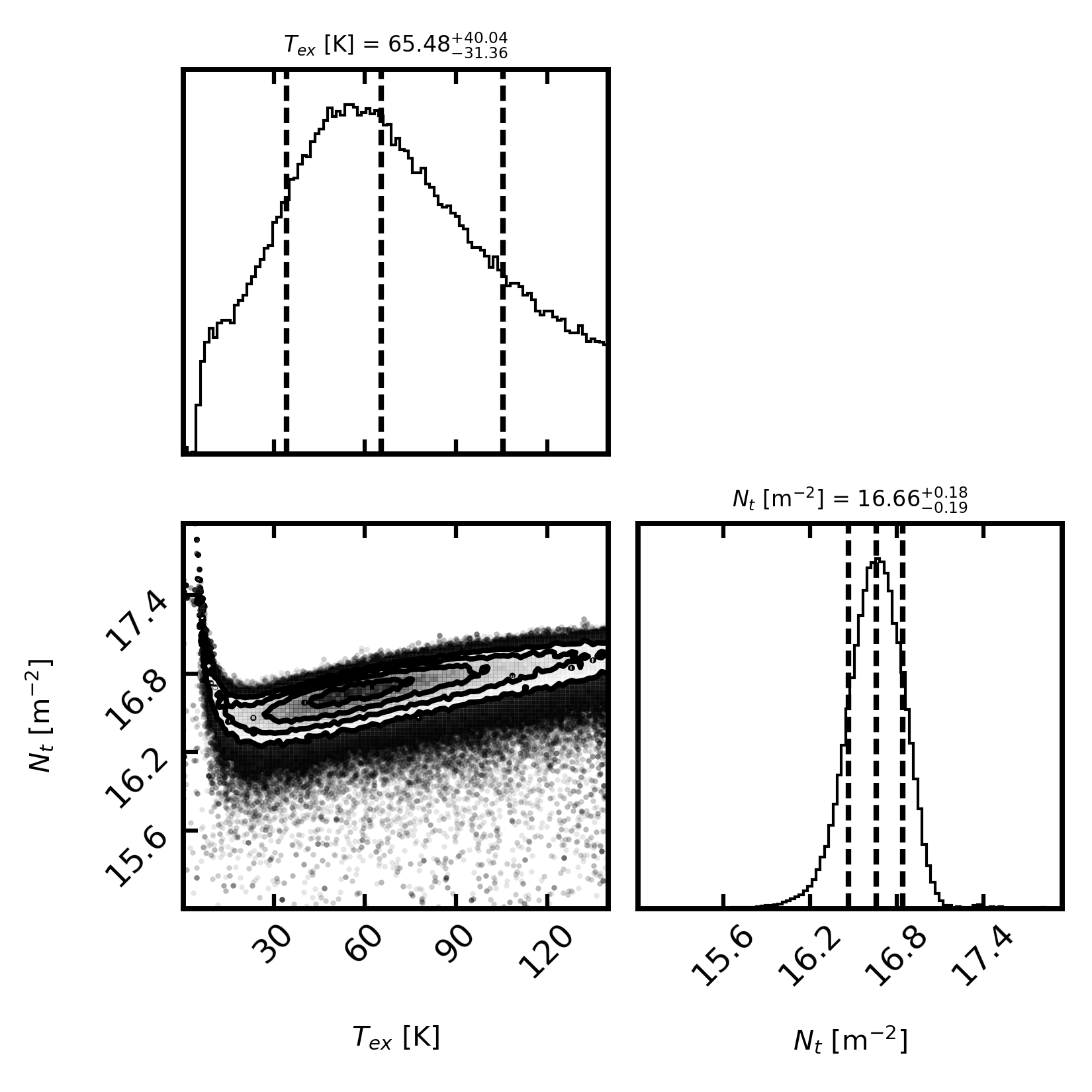}
      \caption{Same as Figure \ref{Fig:corner_aatau} for MWC 758.
              }
         \label{Fig:corner_mwc758}
\end{figure*}

\begin{figure*}
   \centering
   \includegraphics[width=0.48\textwidth]{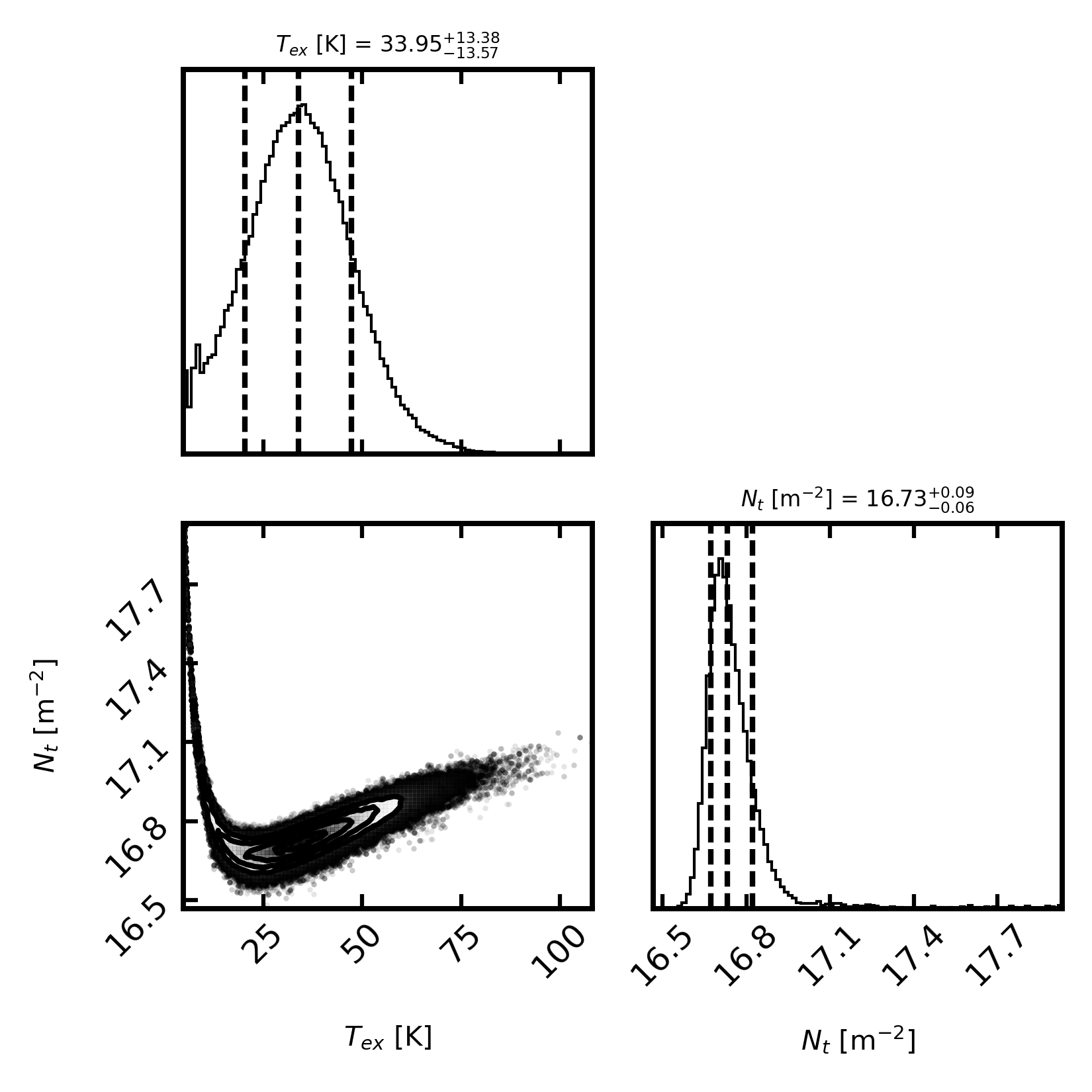}
   \includegraphics[width=0.48\textwidth]{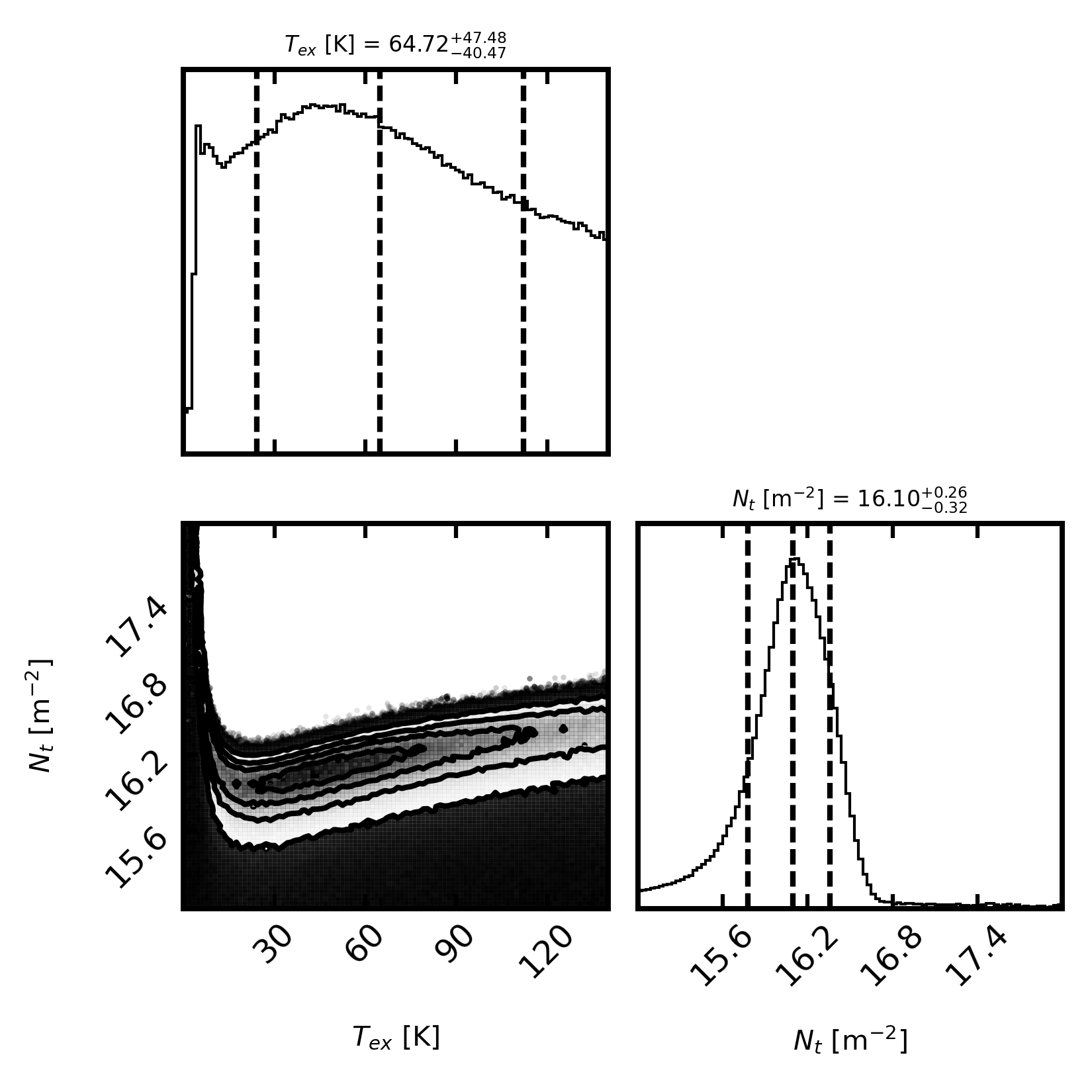}
      \caption{Same as Figure \ref{Fig:corner_aatau} for PDS 66.
              }
         \label{Fig:corner_pds66}
\end{figure*}

\begin{figure*}
   \centering
   \includegraphics[width=0.48\textwidth]{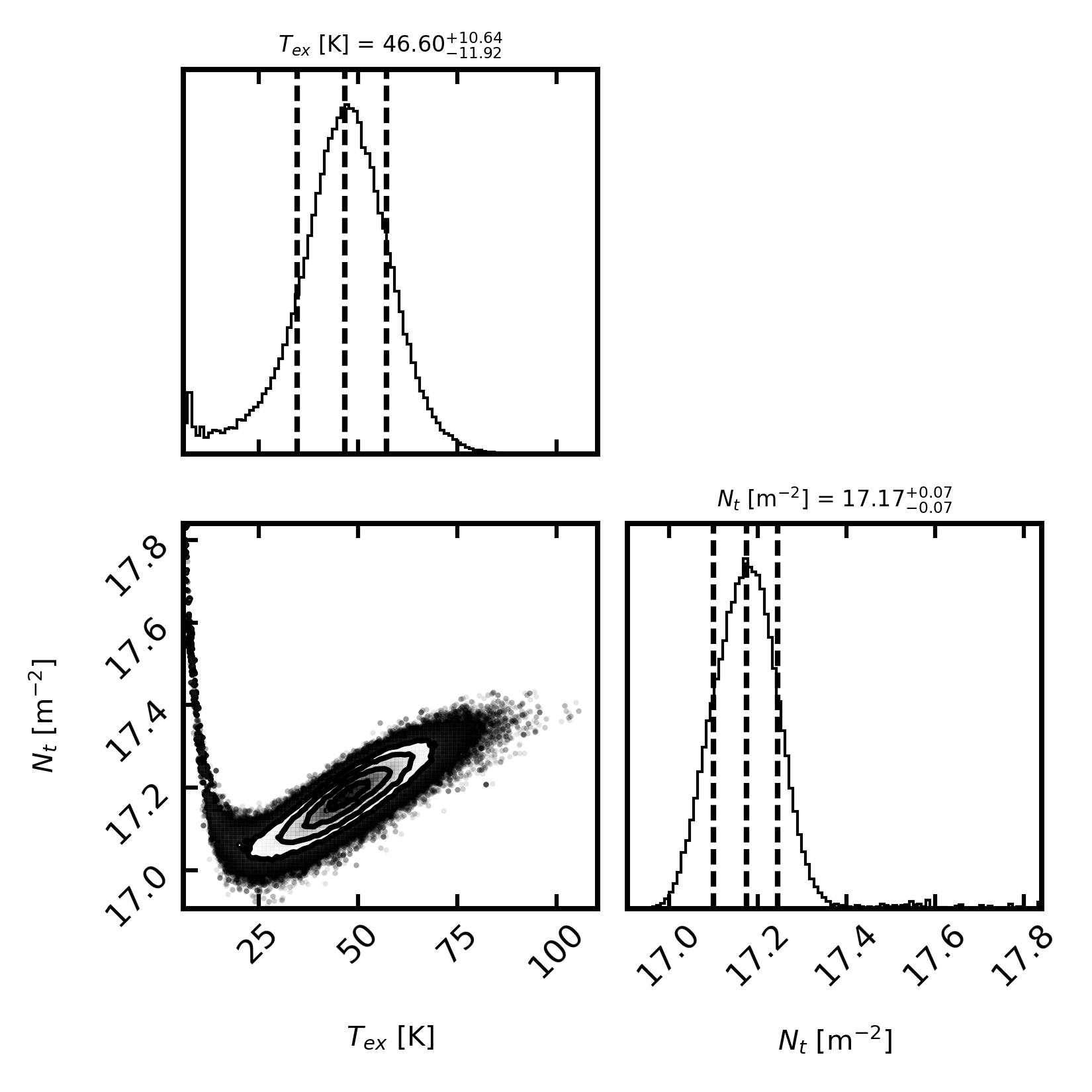}
   \includegraphics[width=0.48\textwidth]{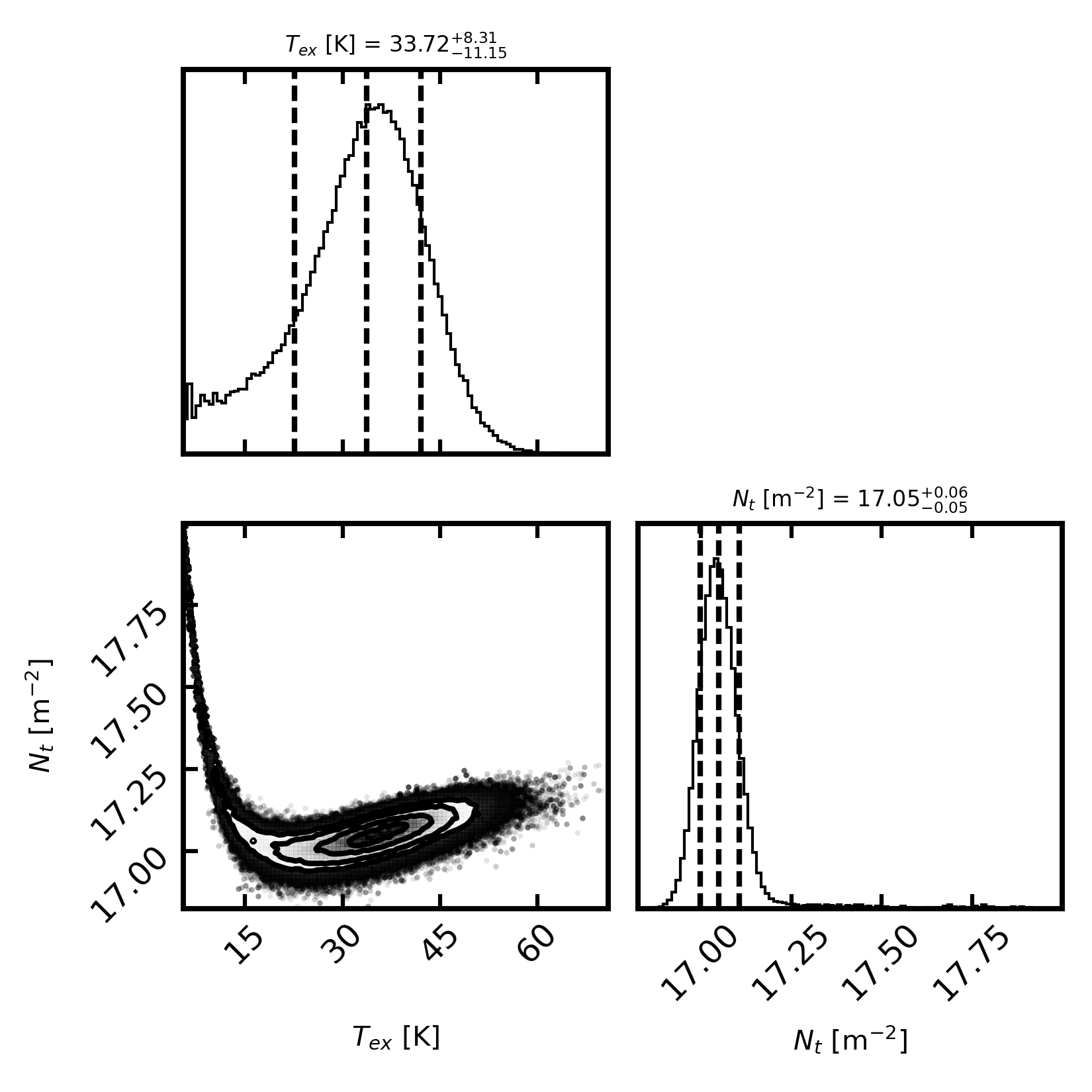}
      \caption{Same as Figure \ref{Fig:corner_aatau} for RX J1615.
              }
         \label{Fig:corner_j1615}
\end{figure*}

\begin{figure*}
   \centering
   \includegraphics[width=0.48\textwidth]{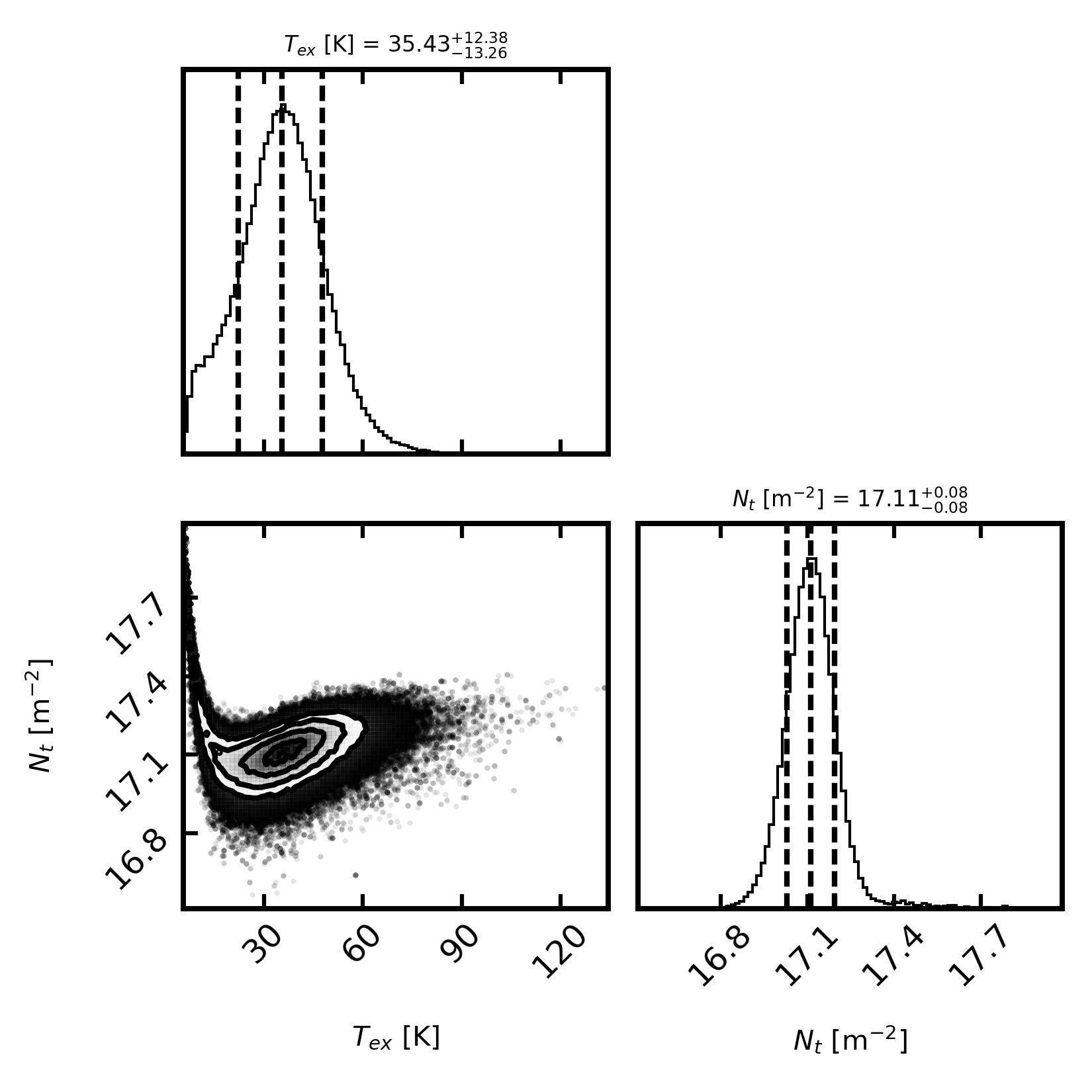}
   \includegraphics[width=0.48\textwidth]{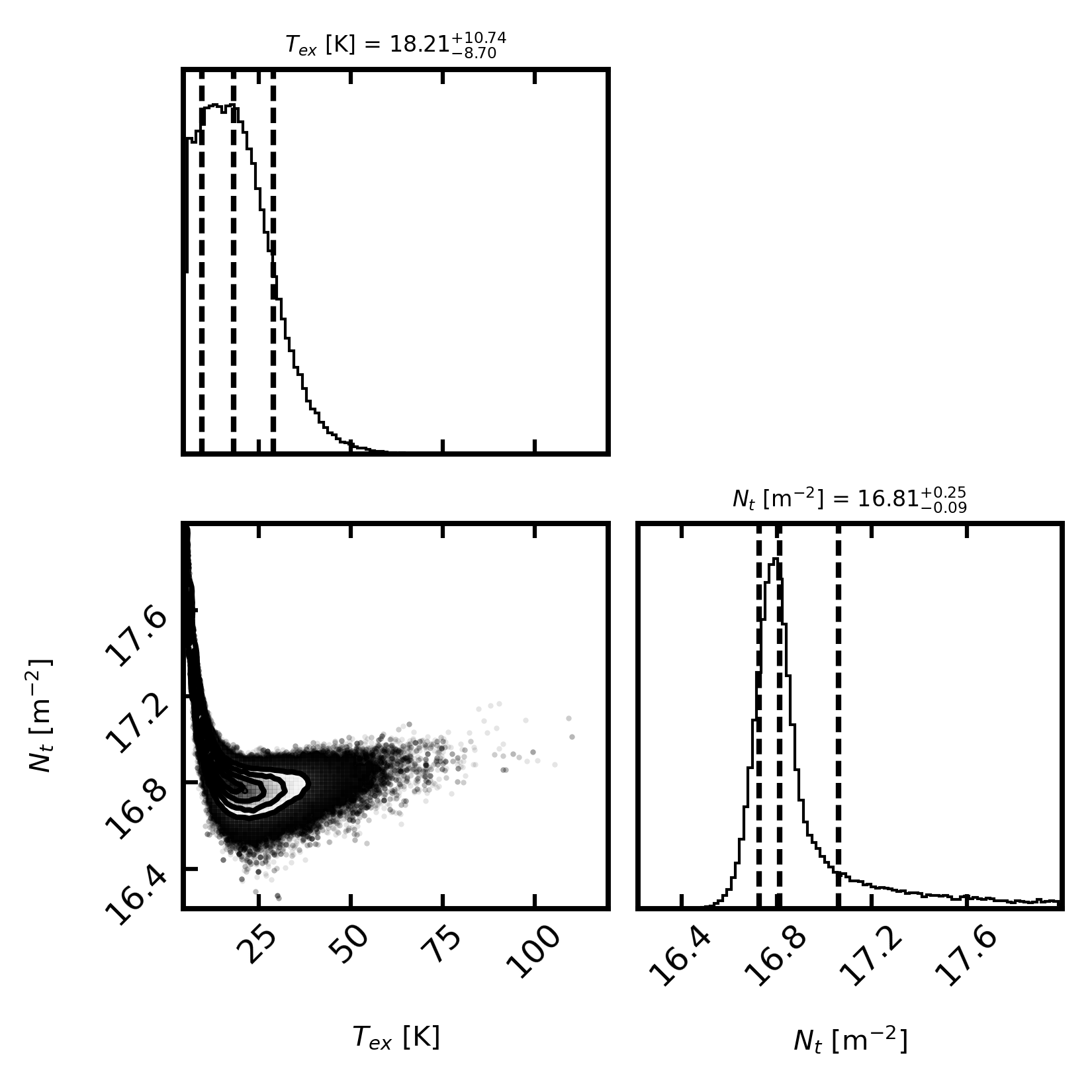}
      \caption{Same as Figure \ref{Fig:corner_aatau} for RX J1842.
              }
         \label{Fig:corner_j1842}
\end{figure*}

\begin{figure*}
   \centering
   \includegraphics[width=0.48\textwidth]{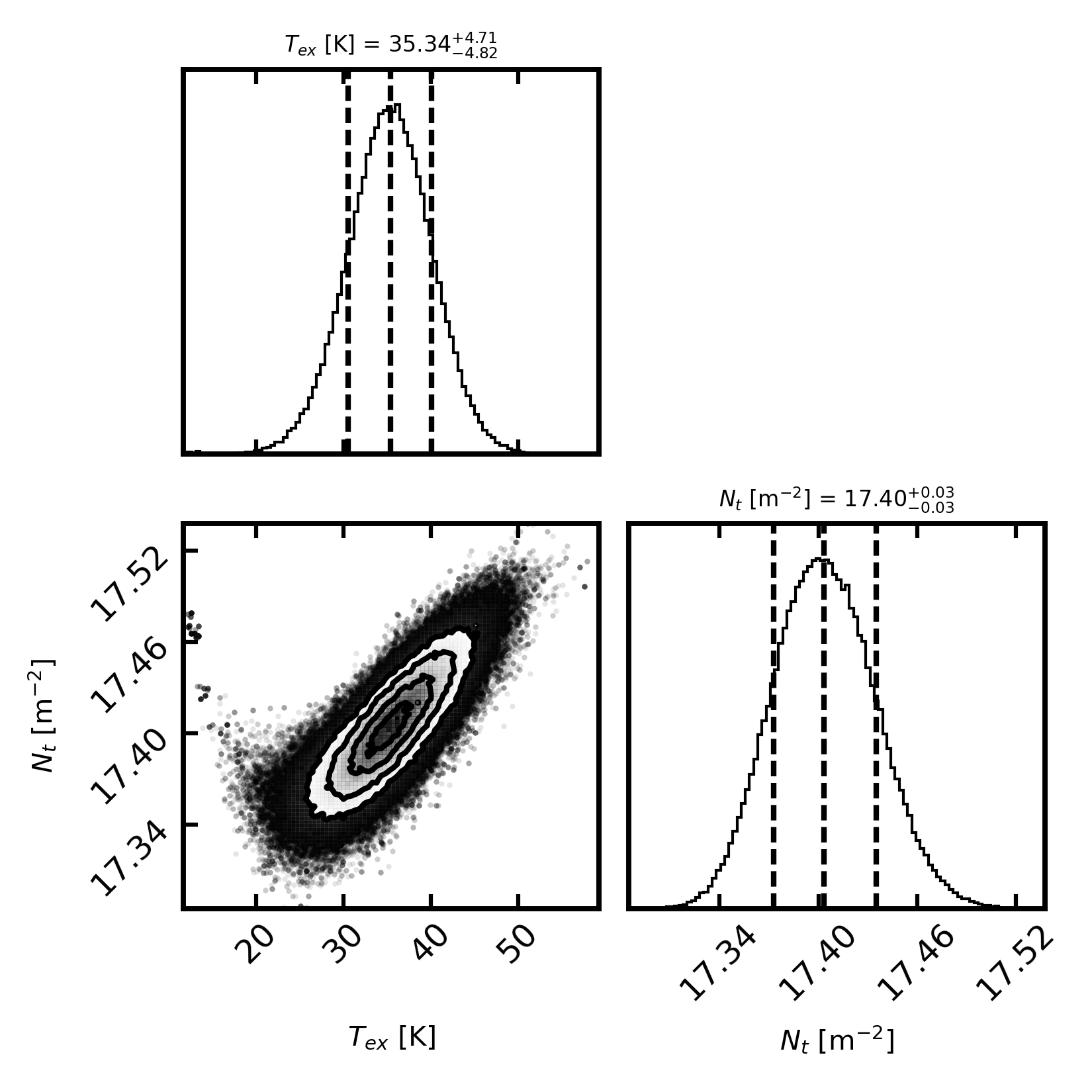}
   \includegraphics[width=0.48\textwidth]{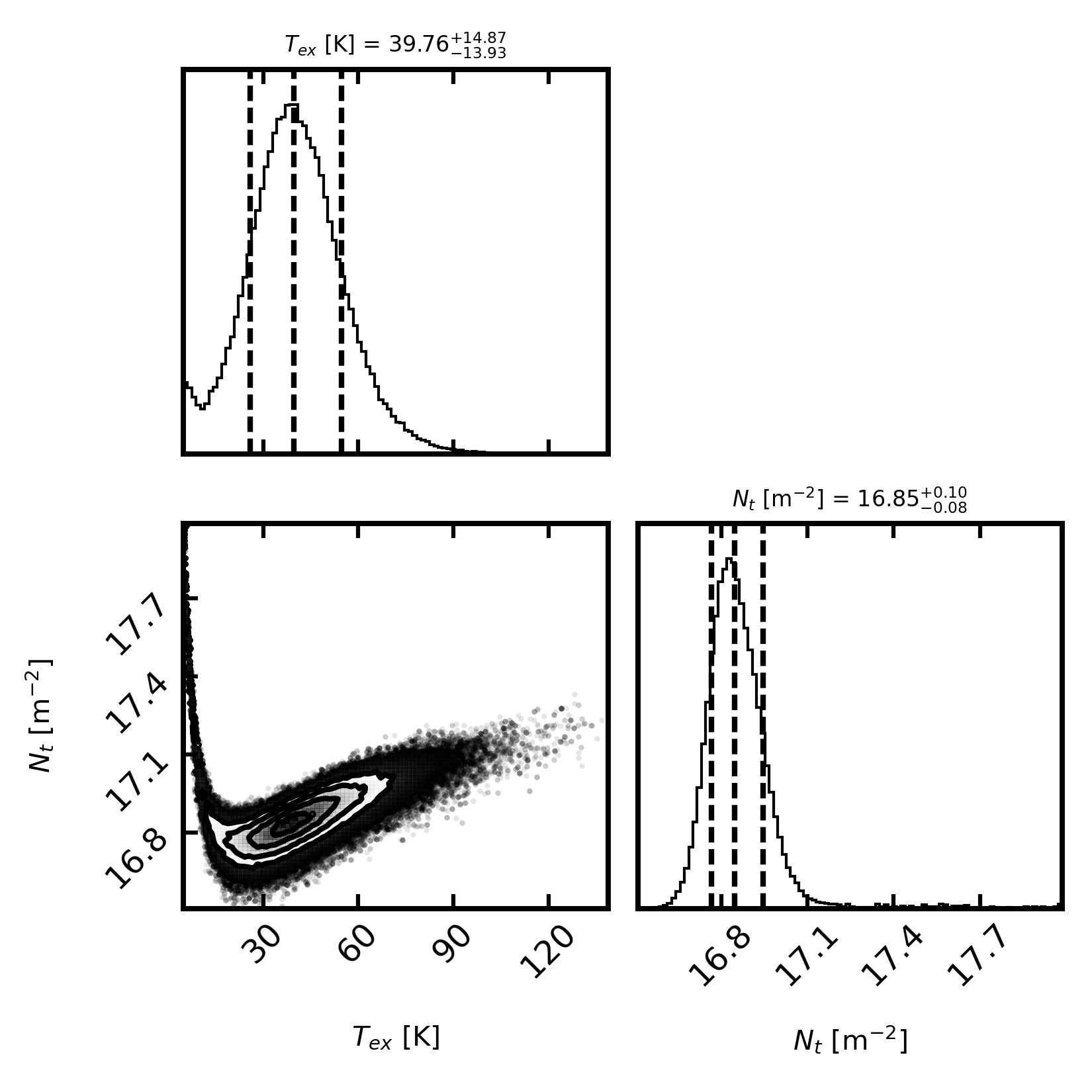}
      \caption{Same as Figure \ref{Fig:corner_aatau} for RX J1852.
              }
         \label{Fig:corner_j1852}
\end{figure*}

\begin{figure*}
   \centering
   \includegraphics[width=0.48\textwidth]{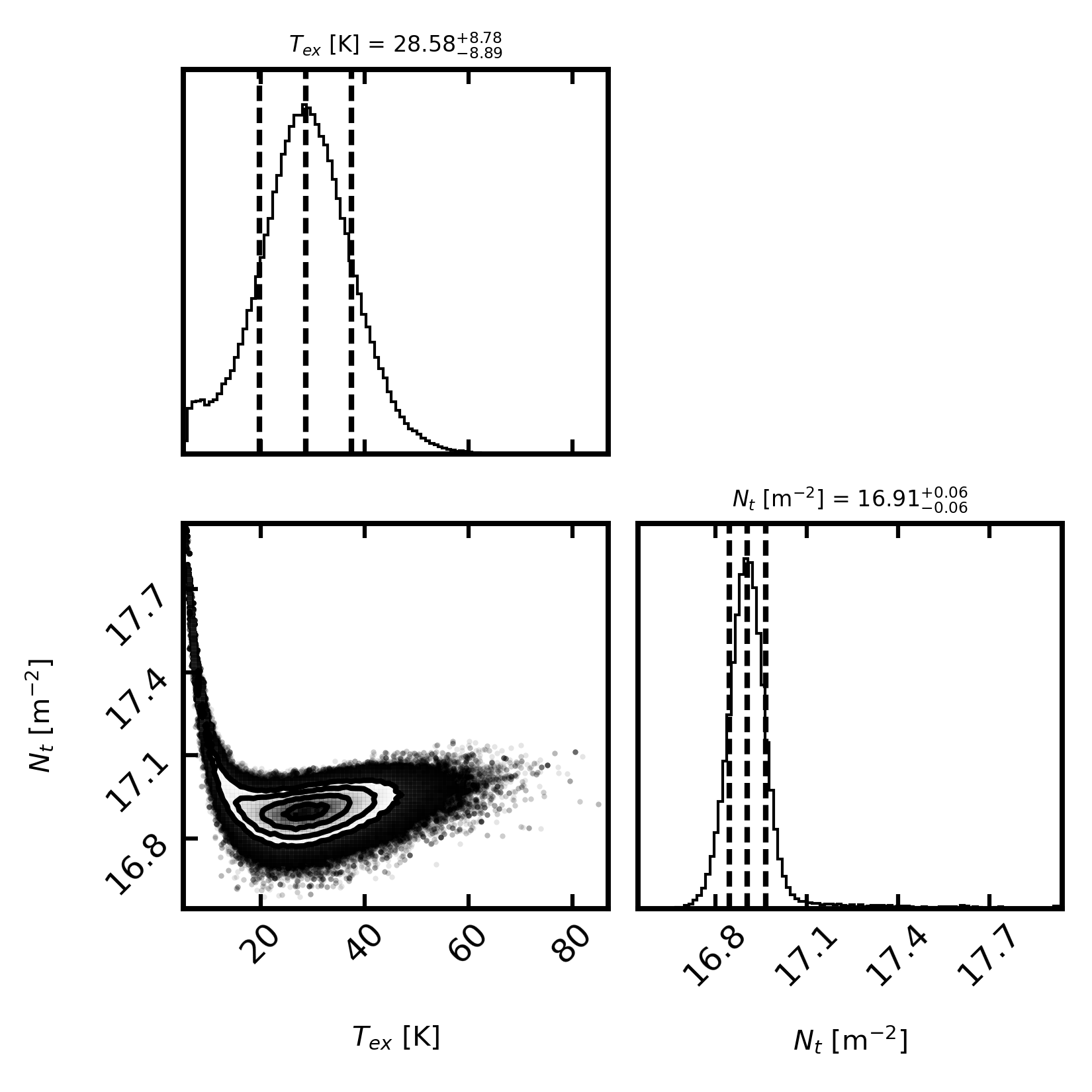}
   \includegraphics[width=0.48\textwidth]{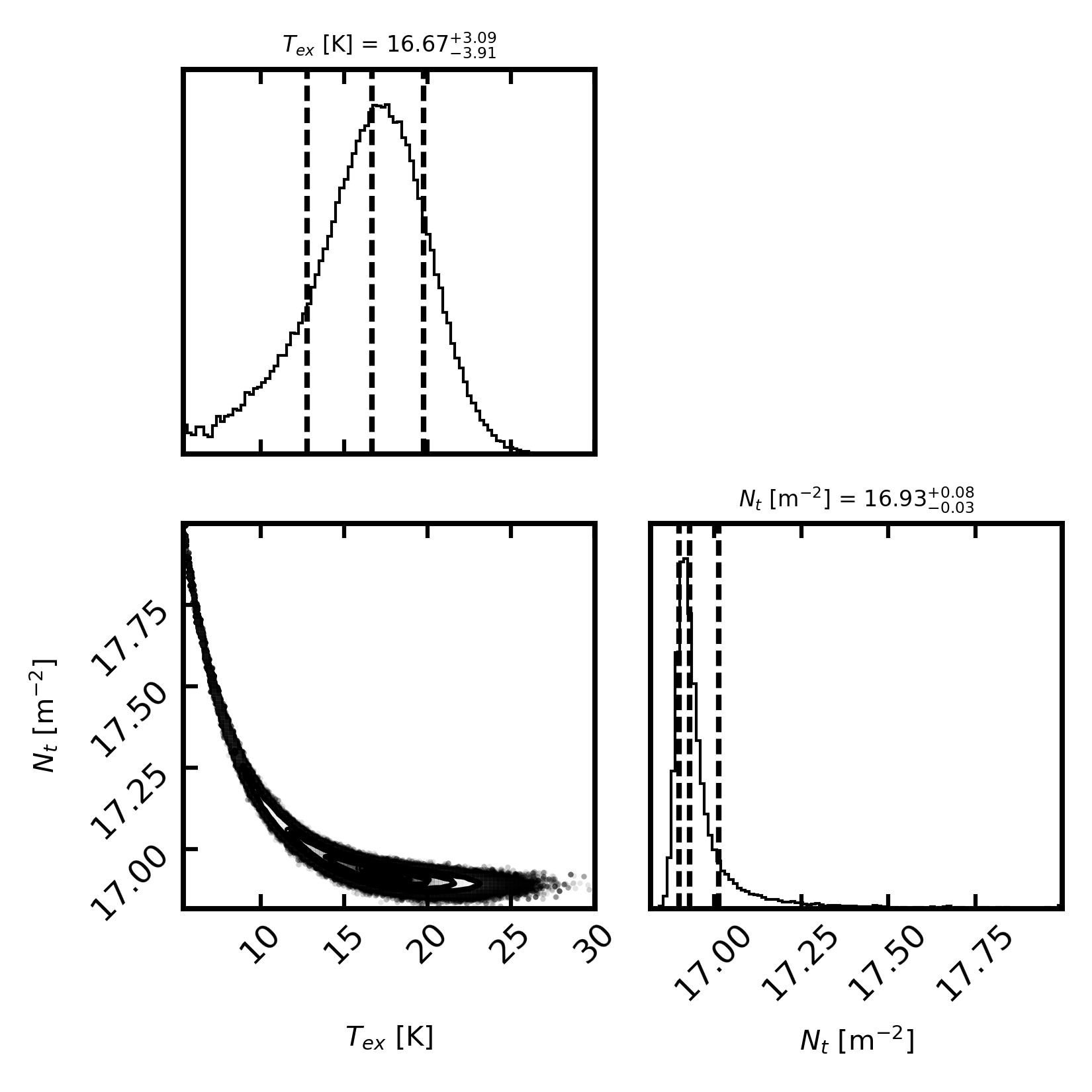}
      \caption{Same as Figure \ref{Fig:corner_aatau} for Elias 2-27.
              }
         \label{Fig:corner_elias}
\end{figure*}

\section{Options for Asymmetry Metrics in Gas Emission}\label{App:NAI}

Establishing a uniform metric to quantify and compare the level of asymmetry of a disk can have multiple intricacies, in particular when considering gas emission. Depending on the type of asymmetry the emission shows and the brightness of the line some asymmetries may be over-weighted, while others may be underestimated or not even considered. We explored three different approaches to use as NAI and benchmark their behavior using different sensitivity thresholds. The analysis on the robustness of the different proposed NAIs in the following subsections endorses our choice presented in Section \ref{Sect: Methods}.

\subsection{Proposed NAIs}

There are different metrics that attempt to quantify the level of asymmetry in the emission, regardless of using integrated intensity or peak intensity maps. We focus our comparison on metrics that quantify point asymmetries, or mirror asymmetries against one of the two natural geometrical axes of the disk.

\subsubsection{Point symmetry}

We can evaluate the symmetry of a disk by comparing each pixel to the opposite location with respect to the disk center. 
It can be defined with the following metric:

\begin{equation}\label{Eq: NAI_PS}
    \mathrm{NAI}_{\mathrm{PS}} = \frac{\sum_{i,j} \vert I_{i,j} - I_{i,j}^{180}\vert }{\sum_{i,j} \vert I_{i,j}\vert},
\end{equation}

\noindent which has been previously used in galactic studies \citep[e.g. ][]{Schade..95, Abraham..96, Davis..2022}. However, this metric is severely affected by the elevated surface of the gas emission. Even in cases where a perfect deprojection was possible, beam smearing effects on the close and far sides of the disk would induce an inclination-dependent trend in the estimated NAI.

\subsubsection{Axial symmetry}

A alternative approach is to assess the asymmetry through reflection of the disk with either its major or minor axis. Unlike point symmetry, this method can effectively capture asymmetries associated with an even number of spiral arms. However, it could cancel out azimuthally elongated features crossing one of the axes- such as horseshoes or dust crescents- which are significant indicators of dynamical activity. Nevertheless, we still evaluate the two different axial symmetries stated above; one with respect to the disk's minor axis (see Eq.. \ref{Eq: NAI}) and the other with respect to its major axis as described below:

\begin{equation}\label{Eq: NAI_MAJ}
    \mathrm{NAI}_{\mathrm{mol}} = \frac{\sum_{i,j} \vert I_{i,j} - I_{i,-j}\vert }{\sum_{i,j} \vert I_{ij}\vert}.
\end{equation}

\noindent The main difference between the two axial symmetries arises from ray-tracing effects along the line of sight to the disk. These effect can introduce artificial asymmetries to the two halves being compared, which are not intrinsic asymmetries but rather observational biases. Such differences become more noticeable at higher inclinations, where the projected overlap between the near and far sides of the disk (especially across the major axis) increases. This can lead to significant overestimation of the intensity's asymmetry. The effect can worsen even more if the emitting layer of the targeted species is elevated in the disk.

\subsubsection{Comparison between different metrics}

\begin{figure*}
   \centering
   \includegraphics[width=0.99\textwidth]{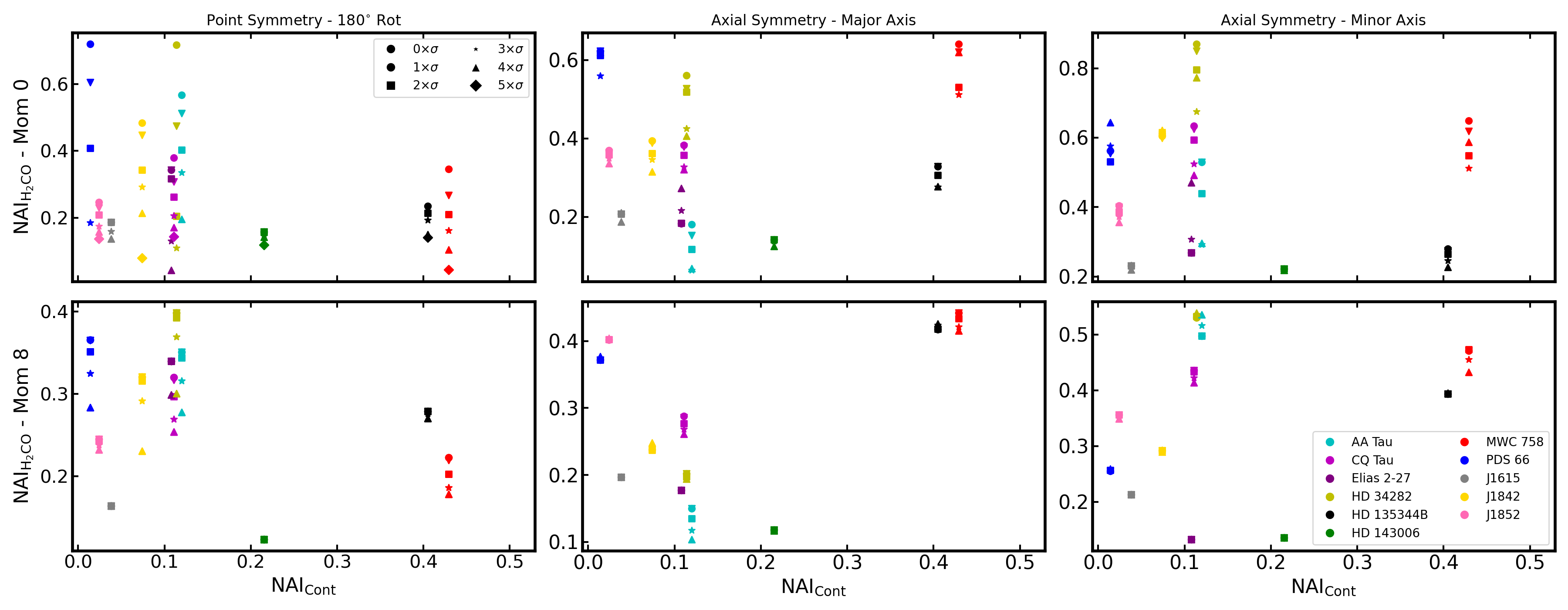}
      \caption{Comparison for the different proposed metrics to assess the asymmetry levels using  integrated intensity (top) or peak intensity (bottom) maps. Different noise thresholds were attempted to evaluate the robustness of the different metrics.
              }
         \label{Fig:nai_comparison}
\end{figure*}

Figure \ref{Fig:nai_comparison} shows the comparison of the different metrics for integrated intensity and peak intensity maps with different noise thresholds. The Figure indicates that the peak intensity maps are more robust than integrated intensity maps as they have a much lower scatter when applying  different sensitivity thresholds. When compared with the axial symmetries, the point symmetry has a much higher scatter for the same disk under different thresholds, making it less reliable to assess the level of symmetry in the gas emission of a particular tracer, in this case H$_2$CO.

\bibliography{references}{}
\bibliographystyle{aasjournal}



\end{document}